\documentclass[twocolumn]{aastex631}
\graphicspath{}
\usepackage{amsmath} 
\usepackage{mathtools}
\usepackage[caption=false]{subfig}

\begin{document}
\title{Lens Modeling of STRIDES Strongly Lensed Quasars using Neural Posterior Estimation}
\author{Sydney Erickson}
\affiliation{Kavli Institute for Particle Astrophysics and Cosmology, Department of Physics, Stanford University}
\affiliation{SLAC National Accelerator Laboratory}
\author{Sebastian Wagner-Carena}
\affiliation{Center for Data Science, New York University}
\affiliation{Center for Computational Astrophysics, Flatiron Institute}
\author{Phil Marshall}
\affiliation{Kavli Institute for Particle Astrophysics and Cosmology, Department of Physics, Stanford University}
\affiliation{SLAC National Accelerator Laboratory}
\author{Martin Millon}
\affiliation{Kavli Institute for Particle Astrophysics and Cosmology, Department of Physics, Stanford University}
\author{Simon Birrer}
\affiliation{Department of Physics and Astronomy, Stony Brook University }
\author{Aaron Roodman}
\affiliation{Kavli Institute for Particle Astrophysics and Cosmology, Department of Physics, Stanford University}
\affiliation{SLAC National Accelerator Laboratory}
\author{Thomas Schmidt}
\affiliation{Department of Physics and Astronomy, University of California, Los Angeles}
\author{Tommaso Treu}
\affiliation{Department of Physics and Astronomy, University of California, Los Angeles}
\author{Stefan Schuldt}
\affiliation{Dipartimento di Fisica, Universita degli Studi di Milano}
\affiliation{NAF -- IASF Milano}
\author{Anowar J.~Shajib}\thanks{NHFP Einstein Fellow}
\affiliation{Kavli Institute for Cosmological Physics, University of Chicago}
\affiliation{Department of Astronomy and Astrophysics, University of Chicago}
\affiliation{Center for Astronomy, Space Science and Astrophysics, Independent University, Bangladesh, Dhaka 1229, Bangladesh}
\author{Padma Venkatraman}
\affiliation{Kavli Institute for Particle Astrophysics and Cosmology, Department of Physics, Stanford University}
\affiliation{Department of Astronomy, University of Illinois at Urbana-Champaign}
\author{The LSST Dark Energy Science Collaboration}

\begin{abstract}

    Strongly lensed quasars can be used to constrain cosmological parameters through time-delay cosmography. Models of the lens masses are a necessary component of this analysis. To enable time-delay cosmography from a sample of $\mathcal{O}(10^3)$ lenses, which will soon become available from surveys like the Rubin Observatory's Legacy Survey of Space and Time (LSST) and the Euclid Wide Survey, we require fast and standardizable modeling techniques. To address this need, we apply neural posterior estimation (NPE) for modeling galaxy-scale strongly lensed quasars from the Strong Lensing Insights into the Dark Energy Survey (STRIDES) sample. NPE brings two advantages: speed and the ability to implicitly marginalize over nuisance parameters. We extend this method by employing sequential NPE to increase precision of mass model posteriors. We then fold individual lens models into a hierarchical Bayesian inference to recover the population distribution of lens mass parameters, accounting for out-of-distribution shift. After verifying our method using simulated analogs of the STRIDES lens sample, we apply our method to 14 \textit{Hubble Space Telescope} single-filter observations. We find the population mean of the power-law elliptical mass distribution slope, $\gamma_{\text{lens}}$, to be $\mathcal{M}_{\gamma_{\text{lens}}}=2.13  \pm 0.06$. Our result represents the first population-level constraint for these systems. This population-level inference from fully automated modeling is an important stepping stone toward cosmological inference with large samples of strongly lensed quasars.

\end{abstract}


\section{Introduction}

 The expansion rate of the Universe today, known as the Hubble constant, $H_0$, is a central point of debate in modern cosmology. To date, there are unresolved discrepancies in the value of $H_0$ derived from early and late Universe probes \citep{Verde_2019, Di_Valentino_2021, Abdalla_2022}. In addition to measuring the current expansion rate, we can measure the expansion rate at past times to build understanding of the nature of dark energy. The expansion history of the Universe is sensitive to any possible evolution of dark energy, often characterized by equation of state parameters ($w_0$,$w_a$). Recent results from the Dark Energy Spectroscopic Instrument collaboration suggest that ($w_0$,$w_a$) may deviate from standard $\Lambda$ cold dark matter values \citep{desicollaboration2024desi}. Considering the ongoing Hubble tension and the possible evidence of the evolution of dark energy, it is important to provide independent measurements of ($H_0$,$w_0$,$w_a$).

Strong gravitational lenses can be used to provide independent constraints on both $H_0$ and ($w_0$,$w_a$) through time-delay cosmography \citep[TDC,][]{refsdal64, Treu_TDC_review, Birrer_TDC_review}. This technique uses lenses of time-variable point sources, such as quasars, for independent single-step distance measurements, without requiring any local calibrations. The Vera C. Rubin Observatory Legacy Survey of Space and Time (LSST) is slated to observe $\mathcal{O}(10^3)$ multiply imaged active galactic nuclei \citep{Oguri_2010}. With this large sample of lenses, TDC can independently constrain the Universe's expansion at percent-level precision \citep{lsst_science_book, Linder_2011, desc_SRD_21, Hogg_2023}.

In addition to cosmography, a large sample of lensed quasars from LSST will be used to enable other scientific studies. Galaxy-scale lensed quasars can be used in combination with galaxy-galaxy lenses to study the mass profiles of elliptical galaxies \citep{shajib2024strong}. Microlensing events can be used to study the physical structure of quasars \citep{vernardos2024microlensing}. Flux ratio anomalies can be used to study dark matter properties \citep{vegetti2024strong}.


For all of these science cases, construction of a lens mass model is crucial. To scale the analysis of lensed quasars for the LSST sample, we need to address the high cost of lens modeling. In a recent lens modeling challenge, a team applying the state-of-the-art forward modeling approach took 500,000 CPU hours and 1700 investigator hours to model 48 lensed quasars \citep{ding2021time}. This translates to roughly 10,400 CPU hours per lens, and 35 investigator hours per lens. This technique will not scale to $\mathcal{O}(10^3)$ time-delay lenses. We, therefore, must invest time in developing fully automated modeling techniques in order to take advantage of data from upcoming large surveys.

\defcitealias{STRIDES}{STRIDES23}

One solution is to further automate the existing modeling process. Several efforts have been made to eliminate investigator intervention and reduce compute time (e.g., \citet{shajib2019every, Shajib_2021, etherington_automated_modeling, STRIDES, ertl2023tdcosmo, Schuldt_2023, tan2023project}). In the analysis by \citet{STRIDES} from the Strong Lensing Insights into the Dark Energy Survey (STRIDES) collaboration (hereafter \citetalias{STRIDES}), CPU time was limited to $<$100 hours per lens, with two lenses in particular (SDSS J0248+1913 and SDSS J1251+2935) taking 11 and 17 CPU hours each. Investigator time was reduced to roughly 10 hours per lens. These techniques are a great improvement in preparing for large samples, but to handle samples with O($10^3$) lenses, we need to reduce investigator time and CPU compute time even further.

Another solution is to use auto-differentiable lensing code to allow for faster inference through gradient informed sampling (e.g., \textsc{GIGALens}; \citep{Gu_2022}; \textsc{herculens}; \citep{Galan_2022}; \textsc{Gravity.jl}; \citep{lombardi_gravity_jl}). In one application of \textsc{herculens} to a cluster-scale lens, it took 140 minutes to produce a posterior for $10^5$ parameters \citep{galan2024el}. Even faster times can be achieved on simpler galaxy-scale lenses. In \citet{lombardi_gravity_jl}, it took under an hour to infer 9 free parameters describing a lensed quasar from 12 observational constraints. These techniques are much faster than traditional CPU-based techniques, but still require explicit choices for nuisance parameters, such as the source light profile and the point spread function (PSF). 

We investigate a third option, which is the use of convolutional neural networks (CNNs) for fully autonomous lens modeling pipelines, as first proposed by \cite{Hezaveh_2017} and \cite{Perreault_Levasseur_2017}. CNN-based techniques have been applied for strong lens modeling in many works (e.g., 
\citet{pearson2019use, madireddy2019modular, holismokes_iv,poh2022strong,holismokes_IX,Legin_2023,Gentile_2023,gawade2024neural}). This technique is advantageous for its fast compute time as well as its implicit marginalization over nuisance parameters.

In this work, we employ CNNs for neural posterior estimation (NPE), a technique that approximates the inference of lens model posterior probability density functions (PDFs) \citep{lueckmann2019likelihoodfree}. A network is trained to predict parameters describing a posterior PDF given an image of a strong lens. This is achieved by using training examples to minimize the distance between network-predicted posteriors and the true posterior. 

NPE has previously been applied for strong lensing modeling. Particularly interesting for TDC analysis is the application of NPE to model simulated \textit{Hubble Space Telescope} (HST) lensed quasars observations for $H_0$ inference \citep{Park_2021}. We aim to build upon this work by introducing the hierarchical framework from \citet{Wagner_Carena_2021} to the analysis of lensed quasars and extending the application to real time-delay lenses.

The inclusion of a hierarchical inference framework serves multiple goals. Firstly, the population model is interesting itself. For example, a constraint on the population distribution of the power-law slope in lens galaxies sheds light on the bulge-halo conspiracy \citep{treu_2006,Buote_2011,Cappellari_2016}. When considering TDC, a hierarchical model of lens properties must be jointly sampled with the cosmological model to properly account for degeneracies \citep{TDCOSMO_IV}. Thus, a key metric of interest for this work is the recovery of the lens parameter population model, which will serve as an important component of future TDC analyses. Finally, if individual lens models are of interest, a population model of lens mass properties can be used to adjust NPE lens models to account for any distribution shift between the training set and test set. This is done through a re-weighting scheme derived in \cite{Wagner_Carena_2021}. In order to apply this correction to lens models, the performance of the hierarchical inference must be verified. We adopt the hierarchical framework developed in \cite{TDCOSMO_IV} and \cite{Wagner_Carena_2021}. To benchmark our progress towards inferring an unbiased population model, we use recovery of the population mean $\mu(\gamma_{\text{lens}})$ as a metric of interest. We choose $\gamma_{\text{lens}}$, because of its interesting astrophysical implications, and because, as demonstrated in \citet{suyu2012cosmography}, error on $\gamma_{\text{lens}}$ directly translates to error on $H_0$ in TDC.

To date, NPE has only been applied to real observations of galaxy-galaxy lenses. Particularly interesting is the application of NPE on Hyper Suprime-Cam (HSC) galaxy-galaxy lenses in \citet{Schuldt_2023} and \citet{gawade2024neural}. In \citet{Schuldt_2023}, predicted time delays from the NPE models are directly compared to time delays computed using mass models from a semi-automated traditional modeling approach. Based on this test, the NPE technique was not recommended for application to HSC time-delay lenses. For the first time, we test the NPE technique on a set of real time-delay lenses, specifically those observed by HST and previously modelled in \citetalias{STRIDES}. We also test an extension of NPE called sequential neural posterior estimation (SNPE), which was first applied to strong lens modeling in \cite{wagnercarena2024strong}. When applying NPE on real lenses, training examples are sampled from a large prior volume, since the range of the prior must be wide enough to include any possible lensing configuration. This large volume results in low density of training samples, which weakens the constraining power of NPE. As shown in \cite{kolmus2024tuning}, the density of training samples that are similar to the test set is a performance driver for NPE. This effect can be alleviated with SNPE, which uses sequential generation of training examples to show the network more informative lenses.  

We aim to answer the following questions:
\begin{itemize}

    \item Will the application of NPE for strong lens mass modeling produce reliable models on real time-delay lenses? In verification tests, what is the percent error per lens on the power-law elliptical mass distribution (PEMD) power-law slope $\gamma_{\text{lens}}$?

    \item How does the application of SNPE compare to NPE? Does the increased sampling density of SNPE improve the precision of lens model posteriors?

    \item What population constraint can we put on the real data using  hierarchical Bayesian inference (HBI)? In verification tests, what is the percent error on the population mean of the PEMD power-law slope $\mu(\gamma_{\text{lens}})$?

\end{itemize}

To answer these questions, we design a series of tests that we run on two distinct test sets. First, we verify our performance on a test set deliberately offset from the prior in the ``shifted" test set. Then, we verify our performance on realistic lensing configurations by creating a test set that closely mimics the data. We call these simulated lenses ``doppelgangers". Finally, we apply our method to the real HST images of STRIDES lenses.   

In Section \ref{section:background}, we introduce our statistical framework for handling many lenses and explain our modeling assumptions. In Section \ref{section:method}, we describe our NPE modeling tool. Then, in Section \ref{section:data}, we describe the datasets we employ our method on. In Section \ref{section:results} we show the results on our verification test sets and the HST data. We end with a discussion of results in Section \ref{section:discussion} and conclusions in Section \ref{section:conclusion}.

\section{Background}
\label{section:background}

We assume a parameterized model for strongly lensed quasars, such that each lens can be described by a set of underlying model parameters, $\xi$. We also assume a population-level model that describes the statistical distributions for the properties of the individual lenses.

\subsection{Individual Lens Models}
\label{section:modeling_choices}

To model a strong gravitational lens, we divide the system into multiple components. We start with the main deflector galaxy. We assign a mass profile and a light profile to this lensing galaxy. Next, we define the background light that sits behind the main deflector along the line of sight. This component includes a source galaxy light profile and a quasar point-source profile. Finally, we include a model for the PSF and additional instrumental effects. 

\subsubsection{Mass Profile}
\label{subsubsection:mass_profile}

Lens mass is defined as the 2D projected surface mass density of the lensing galaxy, also known as the convergence, $\kappa(x,y)$. We assume the lens galaxy mass is described by a PEMD profile \citep{Barkana_1998}:

\begin{equation}
    \kappa(x, y) = \frac{3-\gamma_{\text{lens}}}{2}     \left(\frac{\theta_{E}}{\sqrt{q_{\text{lens}} x^2 + y^2/q_{\text{lens}}}} \right)^{\gamma_{\text{lens}}-1}.
\end{equation}
The power-law slope, $\gamma_{\text{lens}}$, controls the rate at which the density decreases as radius increases. The Einstein radius, $\theta_{\text{E}}$, describes the total amount of mass in the lens. $q_{\text{lens}}$ is the axis ratio, with 1 representing a circle. The coordinates x and y are rotated by angle $\phi_{\text{lens}}$ to be aligned along the major and minor axes, respectively. The coordinates are also shifted to align with the central position of the lens mass, ($x_{\text{lens}}$,$y_{\text{lens}}$). We also include an external shear ($\gamma_{\text{ext}}$, $\phi_{\text{ext}}$) in the model to account for additional distortion from the mass of nearby objects. This shear is described by its strength, $\gamma_{\text{ext}}$, and its orientation angle, $\phi_{\text{ext}}$. 

In total, the mass profile can be described with eight parameters: ($\theta_{\text{E}}$, $\gamma_{\text{ext}}$, $\phi_{\text{ext}}$, $\gamma_{\text{lens}}$, $q_{\text{lens}}$, $\phi_{\text{lens}}$, $x_{\text{lens}}$, $y_{\text{lens}}$). Note that we translate from angular coordinates ($q_{\text{lens}}$, $\phi_{\text{lens}}$), ($\gamma_{\text{ext}}$, $\phi_{\text{ext}}$) to rectangular coordinates ($e_1$, $e_2$), ($\gamma_1$, $\gamma_2$) to avoid the $\pi$ periodicity in $\phi_{\text{lens}}$ and $\phi_{\text{ext}}$. 
\begin{equation}
\begin{split}
    e_1 = \frac{1-q_{\text{lens}}}{1+q_{\text{lens}}}\cos(2\phi_{\text{lens}})
\\
    e_2 = \frac{1-q_{\text{lens}}}{1+q_{\text{lens}}}\sin(2\phi_{\text{lens}})
\end{split}
\end{equation}
\begin{equation}
\begin{split}
    \gamma_1 = \gamma_{\text{ext}}\cos(2\phi_{\text{ext}})
\\
    \gamma_2 = \gamma_{\text{ext}}\sin(2\phi_\text{{ext}}).
\end{split}
\end{equation}

\begin{figure*}[hbt!]
    \centering
    \includegraphics[scale=0.32]{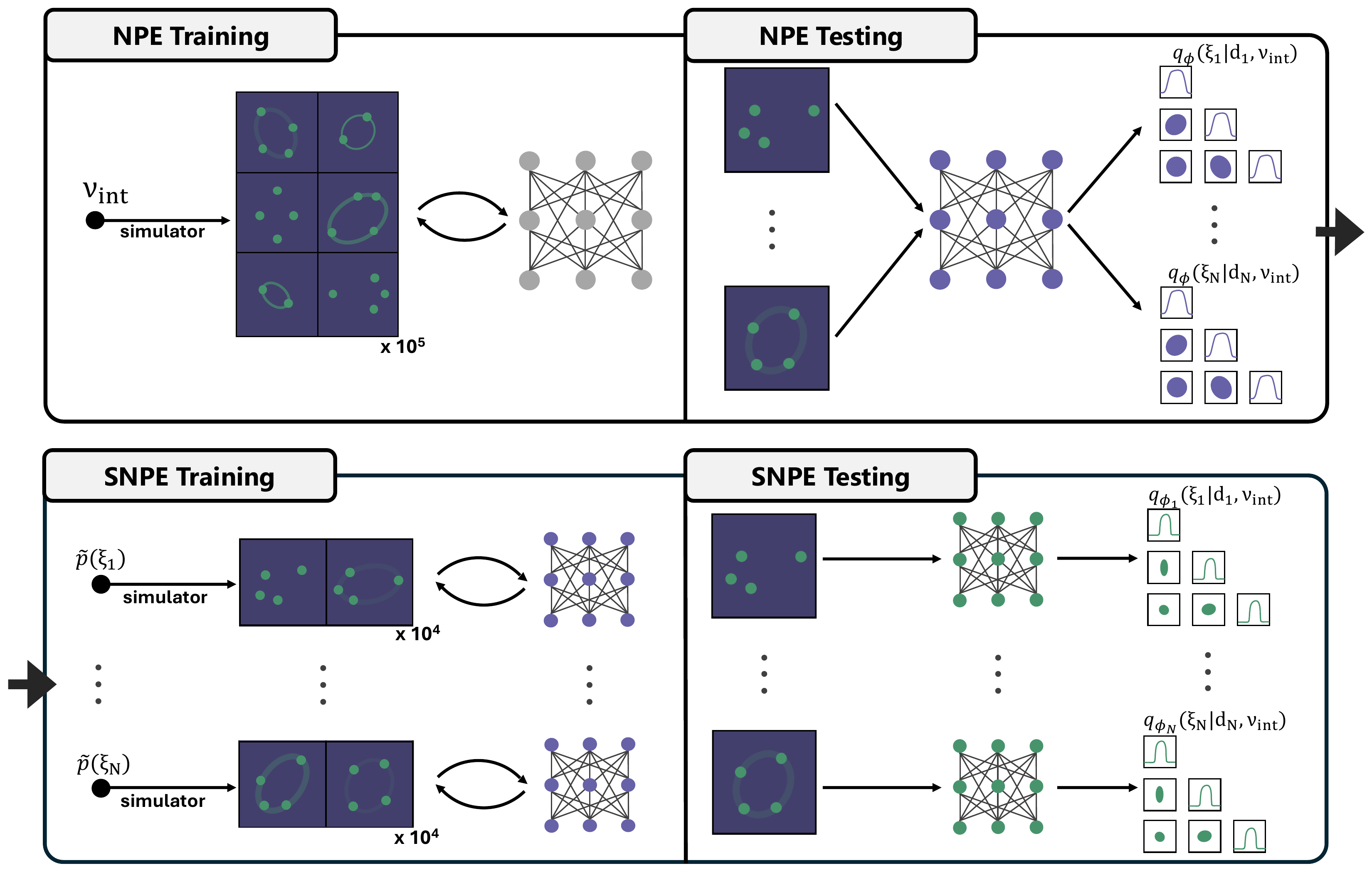}
    \caption{Diagram of the neural posterior estimation (NPE) technique. The first round of training is shown in the top row and described in Section \ref{section:NPE}. 5e5 lenses are sampled from the interim prior, $\nu_{\text{int}}$. Then, a neural network is trained on those examples. At test time, the network takes in an image of a lens, and outputs the parameters describing an approximate lens model posterior $q_\phi(\xi_{k}|d_{k},\nu_{\text{int}})$. The second round of training is sequential, shown in the bottom row and described in Section \ref{section:SNPE}. In this step, 5e4 sequential training examples are generated for each lens in the test set. These new lenses are sampled from a proposal distribution $\tilde{p}(\xi_{k})$ that is informed by the NPE posterior $q_\phi(\xi_{k}|d_{k},\nu_{\text{int}})$. Then, a copy of the neural network is trained for each lens on the sequential training examples. At test time, each lens is passed through its copy of the neural network to produce parameters describing an approximate lens model posterior $q_{\phi_k}(\xi_{k}|d_{k},\nu_{\text{int}})$ }
    \label{fig:method_diagram}
\end{figure*}

\subsubsection{Light Profiles}
\label{subsubsection:light_profile}

We assume a parametric form for the light profile of a galaxy. This profile is used to define light from the source galaxy and the lensing galaxy. We use the elliptical S\'ersic profile \citep{sersic}: 

\begin{equation}
    I(x,y) = I_* \exp \left[ -k_*\left\{ \left( \frac{\sqrt{q_*x^2+y^2/q_*}}{R_*}\right)^{1/n_*} -1\right\}\right].
\end{equation}
The half-light radius, $R_*$, sets the size of the source. The S\'ersic index, $n_*$, determines how concentrated the light profile is. The profile is scaled by surface brightness amplitude $I_*$. Constant $k_*$ depends on $n_*$, and ensures that an ellipse with intermediate-axis length $R_*$ encloses half of the light. The ellipticity is determined by the axis ratio $q_*$. The coordinates x and y are rotated by orientation angle $\phi_*$ to be aligned along the major and minor axis, respectively. The coordinates are also shifted to align with the central position of the galaxy's light ($x_{*}$, $y_{*}$).

The quasar point source is simply defined by a magnitude and a position. We assume the position of the quasar point source coincides with the center of the source galaxy ($x_{\text{src}}$, $y_{\text{src}}$). We additionally account for microlensing by including a random fractional change to the apparent magnitude of each point-source image in the lens plane. This is a simple model of microlensing, primarily included to dis-incentivize the network from using flux ratio information. In future work, more realistic simulations of microlensing could be used to improve the marginalization over this effect. The prior for the fractional microlensing factor, along with all prior choices, is listed in Appendix 
\ref{training_prior}.

\subsubsection{Instrumental Effects}
\label{section:instrumental_effects}

During image simulation, we apply a PSF, which smears out light that hits the telescope. This is especially important for point-source objects, whose appearance is highly affected by the PSF. We use a library of empirically constructed PSF maps for HST's Wide Field Camera 3 (WFC3) UVIS detector that account for variations in focus and location on the chip \citep{stsci_PSFs}\footnote{\url{https://www.stsci.edu/hst/instrumentation/wfc3/data-analysis/psf}}. Our image simulations also include background noise, using a model that accounts for CCD noise, read noise, sky background, and exposure time.

We also model the dithering effect from HST observations to account for the changes it introduces to the effective PSF and noise profile. When dithering, an observation is split into multiple exposures, where each pointing is slightly offset from the others. Then, images are recombined using a drizzling algorithm, which accounts for the offsets. To match the dither strategy of the HST observations from \citetalias{STRIDES}, we generate two dither images offset by 0.5$\arcsec$ in each direction. We then simulate the drizzling effect using the simulation described in \citet{Wagner_Carena_2023}, with a final resolution of 0.04$\arcsec$/pixel, chosen to match the strategy applied in \citetalias{STRIDES}.

\subsection{Population Model for Many Lenses}
\label{subsection:statistical_framework}

%

We aim to infer a population-level  distribution for the properties of individual lensed quasars. We are ultimately interested in a joint inference of the cosmological parameters governing the lenses, $\Omega$, and the distribution of lens mass properties. Joint recovery of these parameters is important when treating time-delay lenses in a hierarchical framework, as shown in \citet{TDCOSMO_IV}. In this work, we focus on constraining the hyperparameters, $\nu$, that govern the distribution of lens model parameters, $\xi$, through the conditional PDF: $p(\xi|\nu)$ (also referred to as the cPDF). We infer a hyperposterior, $p(\nu|\{d\})$ from a dataset of lens observations, $\{d\}$, using HBI. We are especially interested in our ability to recover the hyperparameters that govern the distribution of $\gamma_{\text{lens}}$, since recovery of this parameter is especially important for cosmography \citep{suyu2012cosmography}.

To infer $p(\nu|\{d\})$, we start from Bayes' theorem:
\begin{equation}
    p(\nu | \{d\}) = \frac{p(\{d\}|\nu)p(\nu)}{p(\{d\})}.
\end{equation}
We infer $\nu$ from a population of k lenses. We introduce individual lens parameters $\xi_{k}$, and assume each lens provides an independent constraint. This allows us to break down the problem into constraints on individual lenses.
\begin{equation}
    p(\nu | \{d\}) = p(\nu) \prod_{k} \int \frac{p(d_{k}|\xi_{k})p(\xi_{k}|\nu)}{p(\{d\})} d\xi_{k}
\label{eqn:hi_with_xi_k}
\end{equation}

Note that the population distribution of lens properties inferred from image-based modeling is not completely sufficient to break important degeneracies. Other lens properties must be accounted for, such as the line-of-sight convergence and the internal mass-sheet \citep{Birrer_TDC_review}. These properties do not change image observables, but do change the time delays, in an effect known as mass-sheet degeneracy \citep{falco1985model}. Auxillary data, including galaxy number counts and galaxy kinematics, are included in TDC analyses in order to constrain this effect \citep{TDCOSMO_IV}. When considering lensing degeneracies, the $p(\nu|\{d\})$ inferred from image-based modeling becomes one component of a larger lens population model.

\section{Method}
\label{section:method}

We employ NPE to generate approximations of the lens model posteriors $p(\xi_{k}|d_{k})$. For each lens k, we infer a posterior for PEMD and external shear mass parameters, the lens mass centroid, and the source light centroid: 
\begin{equation}
    \xi_{k} = \{ \theta_{\text{E}}, \gamma_1, \gamma_2, \gamma_{\text{lens}}, e_1, e_2, x_{\text{lens}}, y_{\text{lens}}, x_{\text{src}}, y_{\text{src}} \}.
\end{equation}
These parameters are defined in Section \ref{section:modeling_choices}. We additionally attempt to improve the inference of lens model posteriors by applying SNPE. Then, we infer the parameters governing the population-level model $p(\nu|\{d\})$ from individual lens model posteriors using HBI. 

\subsection{Neural Posterior Estimation}
\label{section:NPE}

Given an image of a strong lens, we can infer a posterior for the lens mass parameters using simulation-based inference. We use NPE, which leverages a neural network to generate approximations of the lens model posteriors. We employ an xResNet-34 CNN to map from input images to a final layer of parameters \citep{he_2016, he_2018bag}. The network architecture was designed for general image-based tasks. Thus, it performs well in many settings, including strong lensing parameter inference \citep{Wagner_Carena_2023}. The mapping is controlled by a set of weights, $\phi$, which are optimized during training. In our setting, the model optimizes the choice of $\phi$ based on the minimization of the NPE loss function, which is evaluated over many pairs of simulated images and their underlying lens model parameters ($d_{k}$,$\xi_{k}$). For a schematic of this method, see the top row of Figure \ref{fig:method_diagram}.

We generate 500,000 examples ($d_{k}$,$\xi_{k}$) for the network to learn from. The data $d_{k}$ is an 80$\times$80 pixel image, simulated to match HST WFC3 observations, as described in Section \ref{section:instrumental_effects}. We define the distribution from which the underlying lensing parameters, $\xi_{k}$, are sampled in Table \ref{tab:training_prior}. This distribution can be thought of as an interim prior, $\nu_{\text{int}}$. The choice of $\nu_{\text{int}}$ must be made carefully. The range of $\nu_{\text{int}}$ is the domain in which the neural network can interpolate. When employing NPE on real data, the range of $\nu_{\text{int}}$ must be wide enough to ensure support for any possible lensing configuration. Additionally, the distribution $\nu_{\text{int}}$ will determine the posterior the network attempts to approximate, particularly if the input data is uninformative. Our choice for $\nu_{\text{int}}$ is described further in Appendix \ref{training_prior}. We generate training samples using the \textsc{paltas}\footnote{\url{https://github.com/swagnercarena/paltas}} package \citep{Wagner_Carena_2023}, which uses \textsc{lenstronomy}\footnote{\url{https://github.com/lenstronomy/lenstronomy}}, a multipurpose strong lensing software program \citep{birrer2018lenstronomy, birrer2021lenstronomy}.

During training, samples from $\nu_{\text{int}}$ are used to optimize the weights of the network's architecture, $\phi$. These weights store a mapping from an input image to a conditional density estimator $q_\phi(\xi_{k}|d_{k},\nu_{\text{int}}$). This estimator approximates the lens model posterior p($\xi_{k}|$$d_{k}$,$\nu_{\text{int}}$). Note that both the conditional density estimator and the posterior are conditioned on the choice of the interim training prior, $\nu_{\text{int}}$. In this work, $q_\phi$ is a 10-dimensional diagonal Gaussian. The assumption of a diagonal covariance matrix is discussed in Section \ref{section:individ_model_functional_form}. To describe $q_\phi$, we need a 10-dimensional $\mu_{k}$ and 10-dimensional $\sigma_{k}$. This requires the network to have a final fully connected layer with 20 outputs. 

To optimize network weights for this task, we minimize the loss function:
\begin{equation}
    L(\phi) = - \sum_{k=1}^N \log \left[ q_\phi(\xi_{k,\text{truth}}|d_{k},\nu_{\text{int}}) \right].
\label{eqn:loss_function}
\end{equation}
 With a sufficiently flexible functional form for $q_\phi$ and a large number of training examples, $N\to \infty$, the approximate posterior converges to the true posterior $q_\phi(\xi_{k}|d_{k},\nu_{\text{int}}) \to p(\xi_{k}|$$d_{k}$,$\nu_{\text{int}})$ when the global minimum of the loss is reached \citep{papamakarios2016fast}. For all aspects of the NPE method, we use the \textsc{paltas} package.  

During training, we minimize the loss function in Equation \ref{eqn:loss_function} using the ADAM optimizer \citep{kingma2014adam}. As an additional augmentation, we randomly rotate the training images each time they are shown to the network. We use 512 images per batch, an initial learning rate of $5\times10^{-4}$, and an exponential decay schedule for the learning rate. To determine when to stop training, we evaluate the loss on a held-out validation set at every epoch. The validation set consists of 5,000 lenses sampled from $\nu_{\text{int}}$. Validation set loss is used as an early stopping criterion. Training is deemed complete when validation loss has not decreased for 10 epochs. With these settings, we reach convergence after 68 epochs, which takes 5.7 hours on an NVIDIA GeForce RTX 2080 Ti GPU. Using the weights from the last epoch, we pass test images through the network to produce approximate posteriors for each lens, $q_\phi(\xi_{k}|d_{k},\nu_{\text{int}}$). 

\begin{figure*}[hbt!]
    \centering
    \includegraphics[scale=0.45]{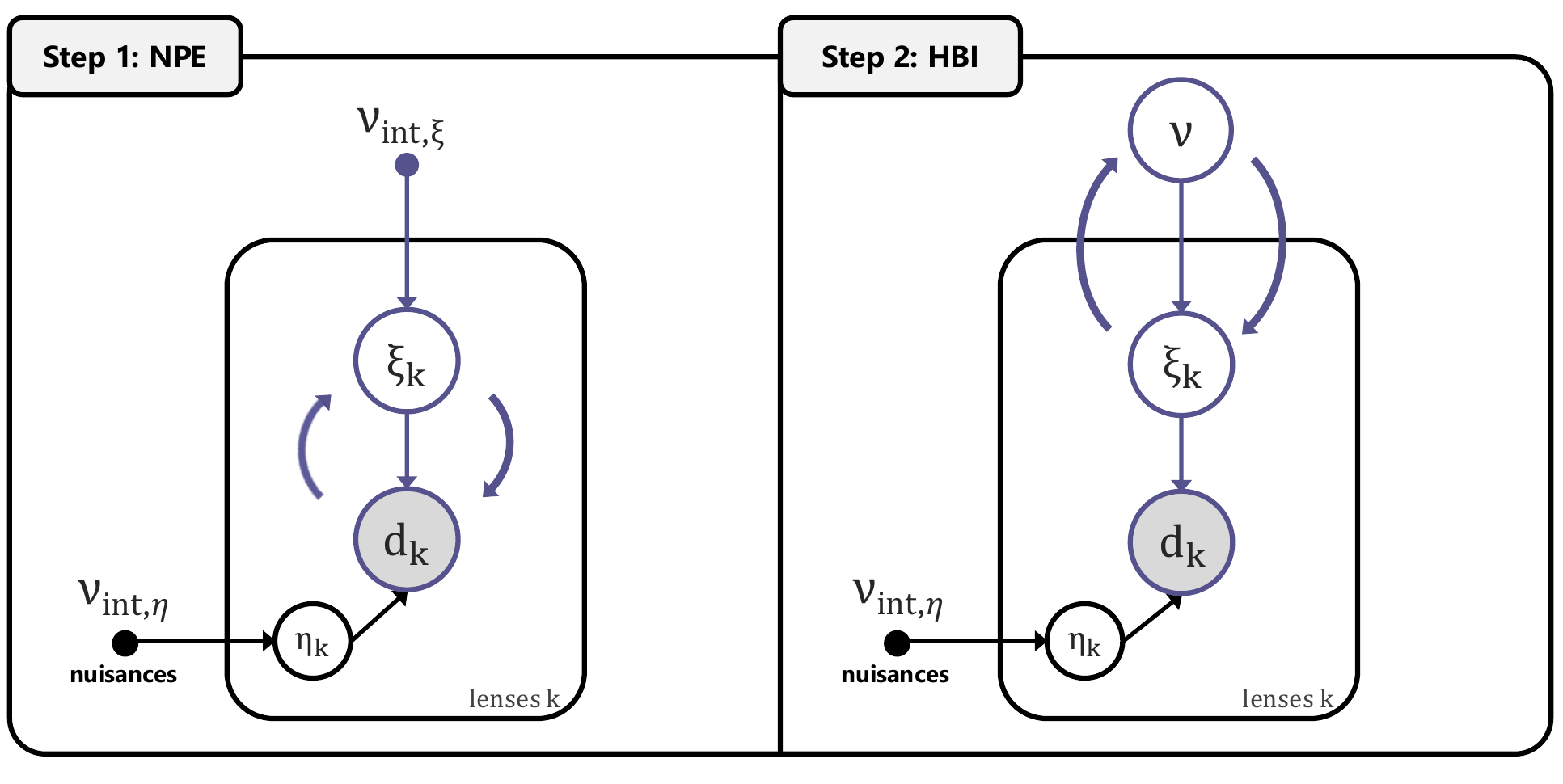}
    \caption{We demonstrate the statistical framework of our hierarchical inference. To generate an image of a strong lens, $d_{\text{k}}$, we need to specify the lens mass parameters, $\xi_{\text{k}}$ and a broader set of nuisance parameters $n_{\text{k}}$. In this work, nuisance parameters include source light, lens light, microlensing effects, PSF kernel, and other instrumental effects. In step 1, NPE, we infer lens model posteriors, $p(\xi_{k}|$$d_{k}$,$\nu_{\text{int}})$, for each lens k. The NPE method allows us to implicitly marginalize over $\eta_{\text{k}}$ at this step. The population-level distribution at this stage is fixed at the training prior, $\nu_{\text{int}}$. In step 2, HBI, we infer a population model, $p(\nu | \{d\})$, over the lens model parameters $\xi$. The population distribution over nuisance parameters, $\nu_{\text{int},\eta}$, remains fixed. }
    \label{fig:pgm_diagram}
\end{figure*}

\subsection{Sequential Neural Posterior Estimation}
\label{section:SNPE}

We investigated multiple ways to further improve the performance of the NPE method. Starting with the simplest investigations, we did not find that longer training or more training examples ($5\times10^{6}$) improved performance. We ultimately hypothesize that the density of training examples near a test example is the most important performance driver, as discussed in \citet{kolmus2024tuning}. While simply using more training examples does improve the density of samples, our parameter space is high-dimensional, which means we may need many orders-of-magnitude more training examples from $\nu_{\text{int}}$ to achieve the density of samples required for significant improvement. To achieve a higher density of samples in a more efficient way, we explore the application of SNPE \citep{papamakarios2016fast}. Succinctly, during NPE training, a majority of time might be spent on examples that are relatively uninformative for the lenses we are interested in. Rather than increase our training sample size, SNPE improves performance by directly modifying our training distribution.

In SNPE, each object in the test set receives a customized training set that is used for an additional round of training. The sequential training examples are generated by a proposal distribution, $\tilde{p}(\xi_{k})$, that is informed by the current NPE posterior. For a schematic of this method, see the bottom row of Figure \ref{fig:method_diagram}.

We employ one sequential step for our SNPE. A proposal distribution $\tilde{p}(\xi_{k})$ is used to generate new samples in a region of interest. Then, these new samples are used to continue training. We use the proposal:
\begin{equation}
    \tilde{p}(\xi_{k}) \propto \left( q_\phi(\xi_{k}|d_{k},\nu_{\text{int}})^n p(\xi_{k}|\nu_{\text{int}})^m \right)^{\frac{1}{m+n}}.
\end{equation}
This proposal allows for a trade-off between exploiting the best guess from the current model and retaining the ability to explore the full parameter space. Larger values of $n$ pull the proposal towards the NPE posterior, while larger values of $m$ hedge the proposal by favoring the original prior. We investigate the choice of $m$ and $n$ in Appendix \ref{appendix:geom_averaging}. For the remainder of the paper, when we refer to the SNPE technique, we are using a proposal with $n=1$ and $m=2$. 

With new training examples in hand, we continue optimization of the network weights, $\phi$. However, we must modify the loss function to account for the change in training distribution. We employ the loss function derived in \cite{greenberg2019automatic}:
\begin{equation}
    L(\phi) = - \sum_{k} \log q_\phi(\xi_{k,\text{truth}}|d_{k},\nu_{\text{int}}) \frac{\tilde{p}(\xi_{k})}{p(\xi_{k}|\nu_{\text{int}})} \frac{1}{Z(d_{k},\phi)}.
\label{eqn:SNPE_loss}
\end{equation}
With this modified loss, we maintain guaranteed convergence of the approximate posterior to the true posterior (see \citealt{greenberg2019automatic} for details). Note that $Z(d_{k},\phi)$ is a normalization constant:
\begin{equation}
    Z(d_{k},\phi) = \int q_\phi(\xi_{k}|d_{k},\nu_{\text{int}}) \frac{\tilde{p}(\xi_{k})}{p(\xi_{k}|\nu_{\text{int}})} d\xi_{k}.
\end{equation}
When $q_\phi$, $\tilde{p}$, and $p$ are all Gaussian, $Z(d_{k},\phi)$ can be computed analytically. 

For each lens in the test set, we generate a copy of the network with randomized weights. Then, 50,000 training examples from proposal $\tilde{p}(\xi_{k})$ are used to optimize the weights of this new network by minimizing Equation \ref{eqn:SNPE_loss}. All training specifications are kept the same as the NPE run, with the learning rate schedule picking up where it left off. Each SNPE training job is run for 10 epochs, which takes roughly 5 minutes on an NVIDIA GeForce RTX 2080 Ti GPU. Using the weights from the 10th epoch, we pass test images through the network to produce approximate posteriors $q_{\phi_{k}}(\xi_{k}|d_{k},\nu_{\text{int}}$). 

\subsection{Hierarchical Bayesian Inference}

After producing mass models for each lens, we are interested in recovering the population-level properties of the lenses. We assume lensing parameters follow a distribution $p(\xi|\nu)$. We constrain $p(\xi|\nu)$ by inferring the values of $\nu$. This results in a hyperposterior, $p(\nu|\{d\})$. We show the statistical framework for our hierarchical inference in Figure \ref{fig:pgm_diagram}.

We are interested in describing the population-level distribution of six lensing parameters: ($\theta_{\text{E}}$,$\gamma_1$,$\gamma_2$,$\gamma_{\text{lens}}$,$e_1$,$e_2$). We assume the population-level distributions for these parameters are Gaussian without any covariance. With this assumption, we describe the population distribution with 12 hyperparameters: six Gaussian means ($\mathcal{M}$) and six standard deviations ($\Sigma$). Note that we use ($\mathcal{M}$,$\Sigma$) to distinguish the Gaussian hyperparameters from the Gaussian individual posterior parameters.

We assume there is no directional preference of the orientation angles of the ellipticity and shear profiles. This equates to enforcing: $\mathcal{M}_{e_1},\ \mathcal{M}_{e_2},\ \mathcal{M}_{\gamma_1},\ \mathcal{M}_{\gamma_2}=0$. We also assume there is no directional preference in the strength of the ellipticity and shear. This equates to enforcing: $\Sigma_{e_1,e_1}=\Sigma_{e_2,e_2}$, and $\Sigma_{\gamma_1,\gamma_1}=\Sigma_{\gamma_2,\gamma_2}$. Enforcing these assumptions reduces $\nu$ to six unique parameters:
\begin{multline}
    \nu = \{\mathcal{M}_{\theta_{\mathrm{E}}}, \mathcal{M}_{\gamma_{\mathrm{lens}}}, \Sigma_{\theta_{\mathrm{E}},\theta_{\mathrm{E}}}, \Sigma_{\gamma_{1/2},\gamma_{1/2}}, \\ \Sigma_{\gamma_{\mathrm{lens}},\gamma_{\mathrm{lens}}}, \Sigma_{e_{1/2},e_{1/2}}\}.
\end{multline}
By combining information from individual lens mass models, we can infer the hyperparameters $\nu$ using hierarchical inference, as described in Section \ref{subsection:statistical_framework}. Note that we infer 10 mass model parameters, but we only include 6 of those parameters in our hierarchical model. Since the individual posteriors are Gaussian, it is trivial to marginalize over the excluded parameters ($x_{\text{lens}}, y_{\text{lens}}, x_{\text{src}}, y_{\text{src}}$). We start from Equation \ref{eqn:hi_with_xi_k}. Then, using Equation C5 from \citet{Wagner_Carena_2021}, we introduce the neural posterior estimate, $q_\phi(\xi_{k} | d_{k}, \nu_{\text{int}})$, to the hierarchical inference:  
\begin{equation}
p(\nu | \{d\}) \propto p(\nu) \prod_{k} \int \frac{q_\phi(\xi_{k} | d_{k}, \nu_{\text{int}}) p(\xi_{k}|\nu)}{p(\xi_{k}|\nu_{\text{int}})} d\xi_{k}.
\label{eqn:HI_final}
\end{equation}

Note that because we explicitly account for the influence of the interim training prior, $\nu_{\text{int}}$, there is a resulting re-weighting 
term, with $p(\xi_{k}|\nu)$ on the numerator, and $p(\xi_{k}|\nu_{\text{int}})$ on the denominator. This is how the hierarchical inference can account for distribution shift between the training prior and the test set. Since we have assumed Gaussian forms for the neural posterior estimate $q_\phi(\xi_{k} | d_{k}, \nu_{\text{int}})$, the training prior $p(\xi_{k}|\nu_{\text{int}})$, and the hyperparameter model $p(\xi_{k}|\nu)$, the integral has an analytic solution.

\begin{figure}[hbt!]
    \centering
    \includegraphics[scale=0.34]{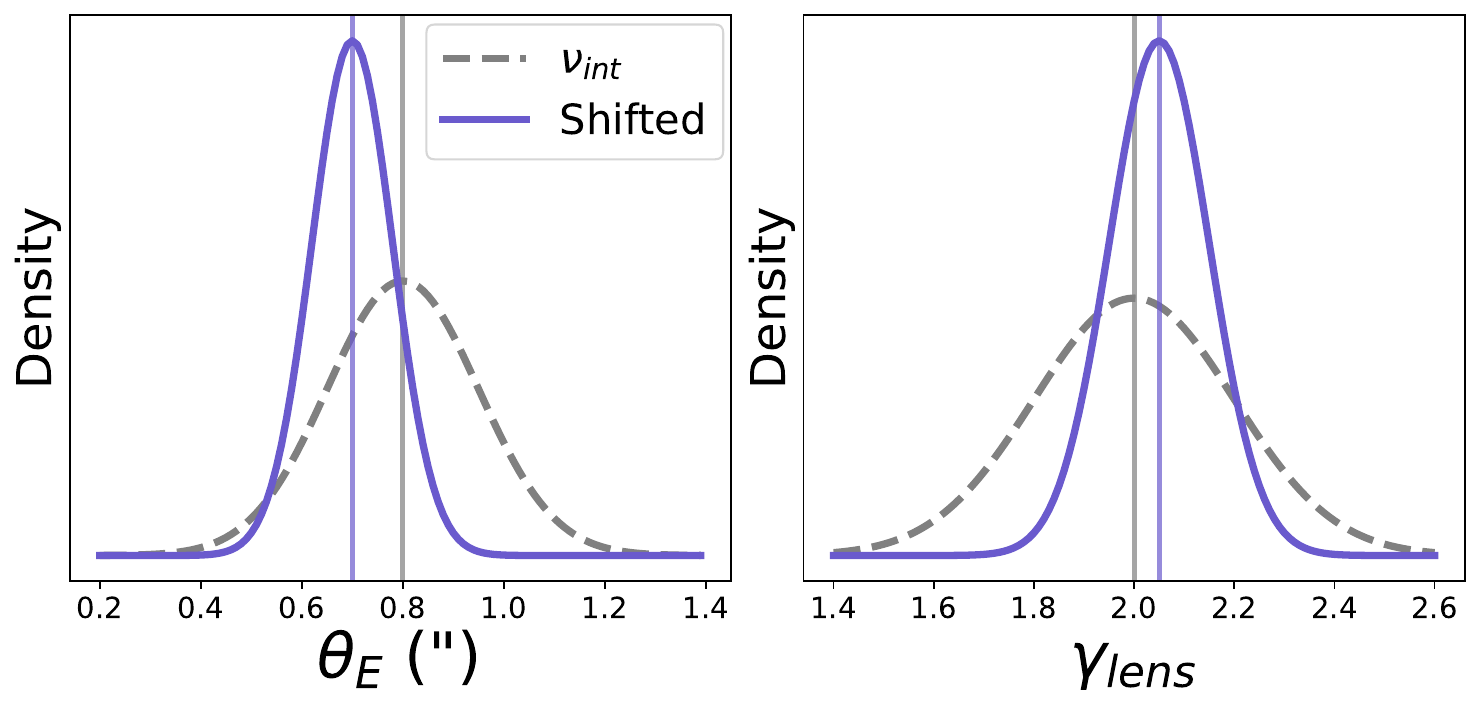}
    \caption{Distribution of the shifted test set (solid purple) compared to the training distribution $\nu_{\text{int}}$ (dashed grey). The shifted test set is designed to test our ability to recover from distribution shift between the training set and the test set. The Gaussian distribution of $\theta_E$ is shifted from $\mathcal{N}$($\mu$=0.8, $\sigma$=0.15) to $\mathcal{N}$($\mu$=0.7, $\sigma$=0.08). The Gaussian distribution of $\gamma_{\text{lens}}$ is shifted from $\mathcal{N}$($\mu$=2.0, $\sigma$=0.2) to $\mathcal{N}$($\mu$=2.05, $\sigma$=0.1).}
    \label{fig:shifted_dists}
\end{figure}

\begin{figure*}[hbt!]
\begin{nolinenumbers}
  \centering
  \subfloat[HST WFC3 images of STRIDES strongly lensed quasars in the F814W band]{\includegraphics[scale=0.6]{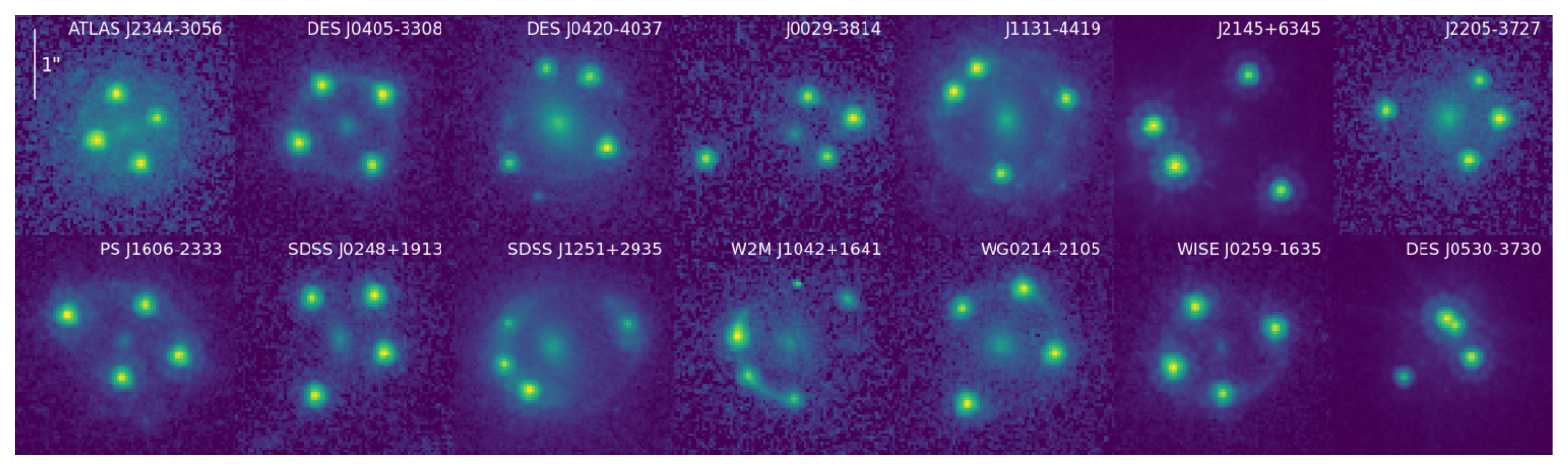}\label{fig:hst_im}}
  \hfill
  \subfloat[Doppelganger simulation counterparts for the STRIDES lenses, created using \textsc{paltas}]{\includegraphics[scale=0.6]{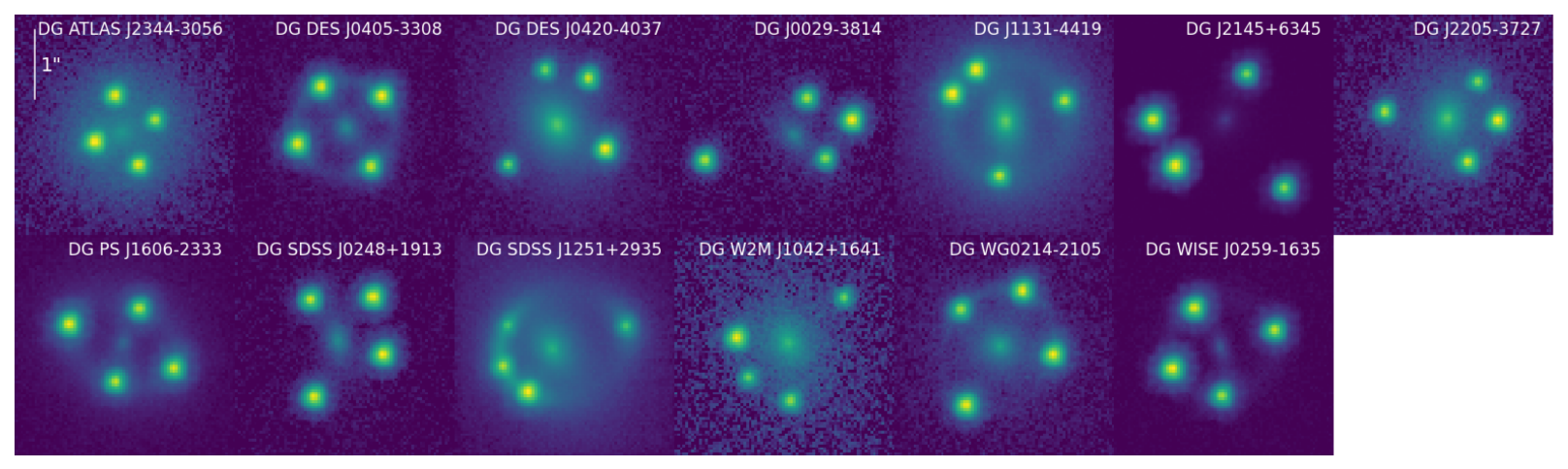}\label{fig:doppel_im}}
  \caption{Comparison of real HST data to the simulated doppelganger test set in the F814W band. Images are 80$\times$80 pixels with 0.04$\arcsec$ resolution. Images are oriented with East to the left, and North to the top. All images are plotted with log-scaled color. Note when using the doppelganger simulation procedure described in Appendix \ref{appendix:doppels}, we were unable to recreate DES J0530$-$3730.}
  \label{fig:hst_and_doppels}
\end{nolinenumbers}
\end{figure*}

We use the posterior in Equation \ref{eqn:HI_final} along with a flat prior to infer $\nu$ using the Markov Chain Monte Carlo ensemble sampler in \textsc{emcee}\footnote{\url{https://github.com/dfm/emcee}} \citep{emcee}. The upper bounds of the prior for population widths ($\Sigma_{\theta_{\mathrm{E}},\theta_{\mathrm{E}}}$, $\Sigma_{\gamma_{1/2},\gamma_{1/2}}$, $\Sigma_{\gamma_{\mathrm{lens}},\gamma_{\mathrm{lens}}}$,$\Sigma_{e_{1/2},e_{1/2}}$) are set to the width of the interim training prior for that parameter (see Table \ref{tab:training_prior}). This is an assumption that the test set distribution is contained within the range of the training prior.

\section{Data}
\label{section:data}

Our primary goal is to apply our method to real data for the first time. To prepare for this application, we first verify the ability of our method to recover valid posteriors. We construct two simulated test sets, the shifted set  (Section \ref{section:shifted_set}) and the doppelganger set (Section \ref{section:doppelganger_set}), for this aim. After verification, our second goal is to assess the robustness of our technique by applying the method to real data. For this aim, we use a set of lensed quasars observed by HST, described in Section \ref{section:hst_set}. As we move from the shifted set, to the doppelganger set, to the real HST data, the datasets become increasingly complex and, therefore, more challenging for our methodology.

\subsection{Shifted Test Set}
\label{section:shifted_set}

The shifted test set is the first of two simulated verification tests. The goal of the shifted test set is to assess our ability to recover from distribution shift. Distribution shift is the difference between the training distribution and the test distribution. Any mismatch may lead to bias in network predictions on the test set, as discussed in Section \ref{section:NPE}. Thus, it is informative to perform a test on a set of lenses with a distribution deliberately offset from $\nu_{\text{int}}$. To create a distribution for the shifted set, we shift and narrow the training distribution in $\theta_E$ and $\gamma_{\text{lens}}$, as shown in Figure \ref{fig:shifted_dists}. We choose these shifts to mimic the shift we see in the doppelganger test set. Twenty samples from the shifted distribution are turned into images using the same simulator as our training set. This test set contains both double and quadruple image configurations. For more details on the creation of the shifted set, see Appendix \ref{appendix:verification_tests}.

\subsection{Doppelganger Test Set}
\label{section:doppelganger_set}

The doppelganger test set is the second simulated verification test. The goal of this test set is to evaluate our performance on images that are as close as possible to real observations. We imitate the STRIDES data by passing the best-fit parameters from \citetalias{STRIDES} forward modeling to our simulator. This way, we have images of realistic lensing configurations that still have a ground truth we can test against. For more details on the doppelganger simulations, see Appendix \ref{appendix:verification_tests}. A comparison of the doppelganger simulations to the HST data is shown in Figure \ref{fig:hst_and_doppels}. 

\subsection{HST Data}
\label{section:hst_set}
We model a set of 14 lensed quasars originally presented and analyzed in \citetalias{STRIDES}. We model only a subset of the original sample since we limited this analysis to images that are contained within an 80$\times$80 pixel cutout. We plan to model the full set in future work. These lenses were observed by programs HST-GO-15320 and HST-GO-15652 (PI: Treu). We model images taken by HST's WFC3 in the F814W band, using data products derived in \citetalias{STRIDES}. We refer the reader to \citetalias{STRIDES} for more details on the individual lenses and data reduction. A gallery of the observations is shown in Figure \ref{fig:hst_im}.

\section{Results}
\label{section:results}

As introduced in Section \ref{section:data}, we design two verification tests to establish confidence in our method on labelled examples. We present our results on these tests first. Then, we run our method on real data for the first time, giving us an opportunity to assess our robustness.

First, we review the outputs of each modeling step and introduce the metrics used to evaluate these outputs in Sections \ref{subsection:metrics} and \ref{subsection:hbi_metrics}. Then, we show the performance of our method on the verification test sets in Section \ref{subsection:verification_results}. Finally, we present our results on the real HST lenses in Section \ref{subsection:data_results}. 

\subsection{NPE Modeling Metrics}
\label{subsection:metrics}

The first step of the modeling process is neural posterior estimation of lens mass models, which produces approximate posteriors $q_\phi(\xi_{k}|d_{k},\nu_{\text{int}})$ for each lens k. To evaluate our initial mass modeling step, we want to evaluate how well $q_\phi(\xi_{k}|d_{k},\nu_{\text{int}})$ captures the ground truth $\xi_{\text{k,truth}}$ on our verification tests. 

First we check for introduction of bias. Bias manifests as systematically high or low prediction. To check for bias, we compute the median of the difference between the mean of the posterior, $\mu_{k}$ and the ground truth, $\xi_{\text{k,truth}}$, across the test set:
\begin{equation}
    \text{ME} = \text{median}_{k} \left\{\mu_{k} - \xi_{k,\text{truth}} \right\}
\label{eqn:ME}.
\end{equation}
Because our test sets contain a small number of lenses -- 20 in the shifted set and 13 in the doppelganger set --a deviation from zero is indicative of but does not guarantee bias.

\begin{figure*}[hbt!]
\begin{nolinenumbers}
\begin{center}
\subfloat[Shifted test set]{\includegraphics[scale=0.35]{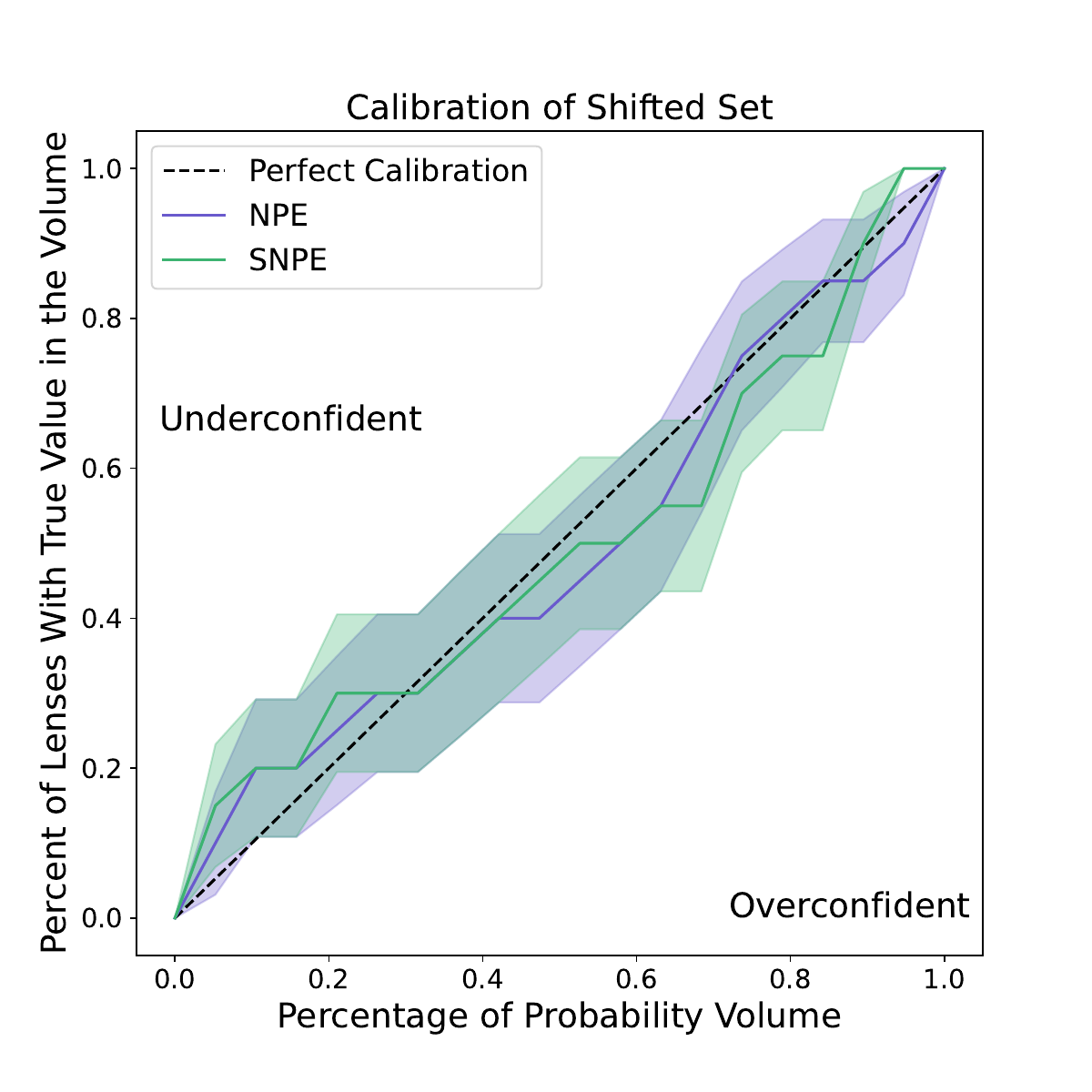}\label{shifted_calib}}
\quad
\subfloat[Doppelganger test set]{\includegraphics[scale=0.35]{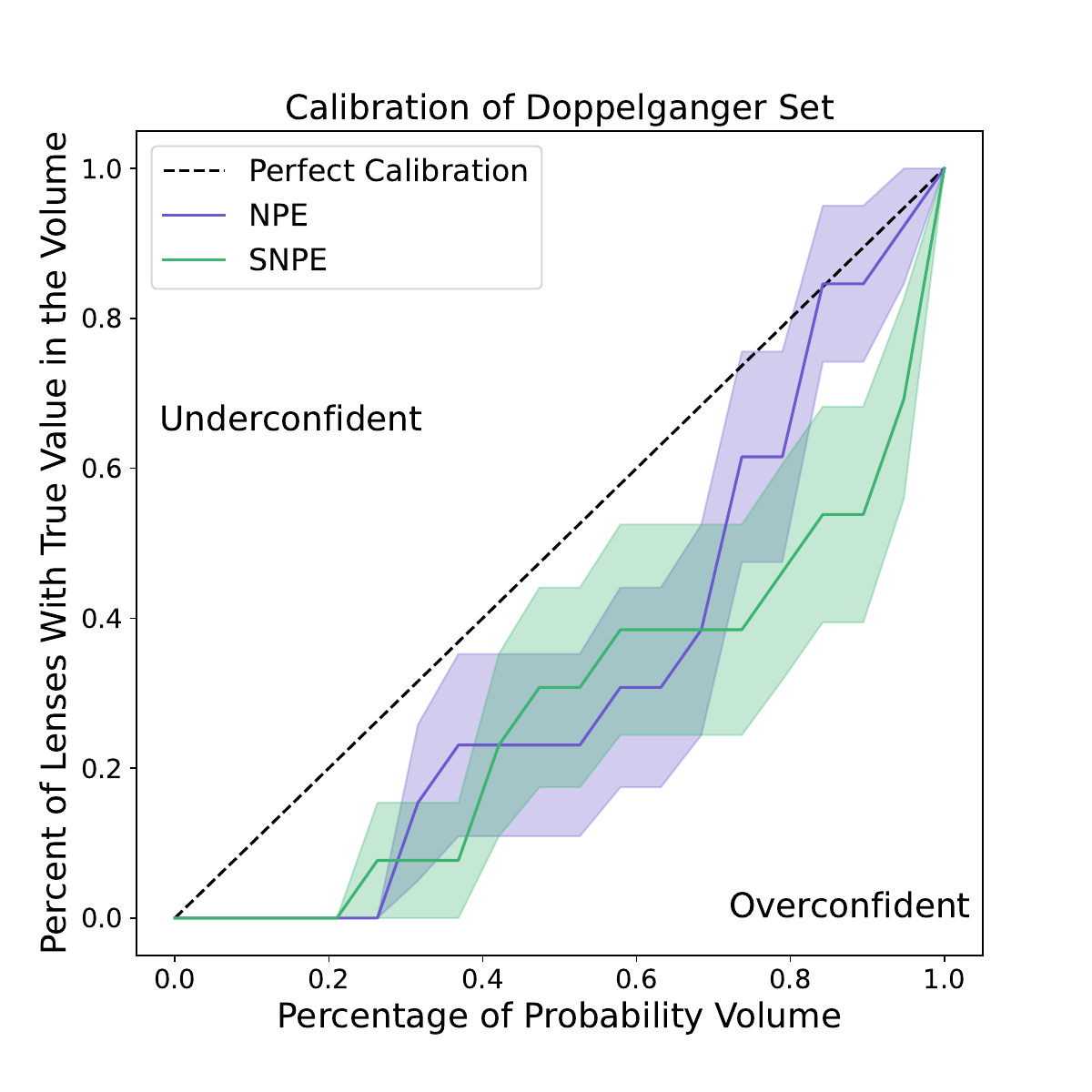}\label{doppel_calib}}
\caption{Calibration curves for the verification test sets. In perfectly calibrated posteriors (dashed line), a given x\% of the probability volume contains the truth x\% of the time. Calibration of NPE posteriors is shown in purple. Calibration of SNPE posteriors is shown in green. The shaded region encompasses 1$\sigma$ uncertainty. Doppelganger posteriors are more overconfident than shifted set posteriors for both modeling methods, which is discussed in Section \ref{section:shifted_vs_doppels}. }
\label{calibration_curves}
\end{center}
\end{nolinenumbers}
\end{figure*}

\begin{table*}
\begin{nolinenumbers}
\centering
\begin{tabular}{| c  c | c | c | c | c | c | c | c | c | c | c |}
    \hline
    & & $\theta_\mathrm{E}$ ($\arcsec$) & $\gamma_1$ & $\gamma_1$ & $\gamma_\mathrm{lens}$ & $e_1$ & $e_2$ & $x_{\text{lens}}$ ($\arcsec$) & $y_{\text{lens}}$ ($\arcsec$) & $x_{\text{src}}$ ($\arcsec$) & $y_{\text{src}}$ ($\arcsec$)\\
    \hline
    \textbf{ME} & NPE & -0.0 & 0.0 & -0.0 & 0.01 & 0.0 & -0.02 & -0.0 & -0.0 & 0.0 & 0.0 \\
    & SNPE & -0.0 & 0.0 & 0.0 & 0.03 & 0.01 & -0.02 & -0.0 & 0.0 & -0.0 & 0.0 \\
    \hline
    \textbf{MAE} & NPE &0.0 & 0.01 & 0.01 & 0.09 & 0.03 & 0.03 & 0.0 & 0.0 & 0.01 & 0.0 \\

    & SNPE & 0.01 & 0.01 & 0.01 & 0.09 & 0.03 & 0.02 & 0.0 & 0.0 & 0.01 & 0.01 \\
    \hline
    \textbf{M($\sigma$)} & NPE & 0.01 & 0.02 & 0.02 & 0.12 & 0.04 & 0.04 & 0.0 & 0.0 & 0.01 & 0.01\\

    & SNPE &0.01 & 0.02 & 0.02 & 0.1 & 0.04 & 0.04 & 0.01 & 0.01 & 0.01 & 0.01\\
    \hline
\end{tabular}
\caption{Metrics for \textbf{shifted set} lens models. We report median error (ME, Eqn. \ref{eqn:ME}), median absolute error (MAE, Eqn. \ref{eqn:MAE}), and median $\sigma$ (M($\sigma$), Eqn. \ref{eqn:MS}). We report performance for both NPE and SNPE modeling.}
\label{tab:metrics_shifted}
\end{nolinenumbers}
\end{table*}

\begin{table*}
\begin{nolinenumbers}
\centering
\begin{tabular}{| c  c | c | c | c | c | c | c | c | c | c | c |}
    \hline
    & & $\theta_\mathrm{E}$ ($\arcsec$) & $\gamma_1$ & $\gamma_1$ & $\gamma_\mathrm{lens}$ & $e_1$ & $e_2$ & $x_{\text{lens}}$ ($\arcsec$) & $y_{\text{lens}}$ ($\arcsec$) & $x_{\text{src}}$ ($\arcsec$) & $y_{\text{src}}$ ($\arcsec$)\\
    \hline
    \textbf{ME} & NPE & -0.0 & 0.02 & 0.0 & 0.09 & 0.01 & 0.01 & 0.0 & -0.01 & 0.0 & -0.0 \\
    & SNPE & -0.0 & 0.01 & 0.0 & 0.09 & 0.02 & 0.02 & 0.0 & -0.0 & 0.0 & -0.0 \\
    \hline
    \textbf{MAE} & NPE & 0.01 & 0.02 & 0.02 & 0.12 & 0.02 & 0.04 & 0.01 & 0.01 & 0.01 & 0.01 \\

    & SNPE & 0.01 & 0.02 & 0.01 & 0.1 & 0.03 & 0.02 & 0.01 & 0.0 & 0.0 & 0.0 \\
    \hline
    \textbf{M$(\sigma$)} & NPE & 0.0 & 0.02 & 0.02 & 0.16 & 0.03 & 0.03 & 0.0 & 0.0 & 0.01 & 0.01\\

    & SNPE & 0.0 & 0.01 & 0.01 & 0.12 & 0.03 & 0.03 & 0.0 & 0.0 & 0.01 & 0.0 \\
    \hline
\end{tabular}
\caption{Metrics for \textbf{doppelganger set} lens models. We report median error (ME, Eqn. \ref{eqn:ME}), median absolute error (MAE, Eqn. \ref{eqn:MAE}), and median $\sigma$ (M($\sigma$), Eqn. \ref{eqn:MS}). We report performance for both NPE and SNPE modeling.}
\label{tab:metrics_doppel}
\end{nolinenumbers}
\end{table*}

Next, we evaluate the accuracy of the network's predictions. Accuracy captures how close the prediction is to the ground truth. A modeling technique could achieve zero bias without actually learning information by simply predicting the mean of the test distribution. Evaluating our accuracy ensures that the network is learning information from the data. To benchmark accuracy, we use median absolute error (MAE), which is also evaluated on the mean of the posterior: 
\begin{equation}
    \text{MAE} = \text{median}_{k} \left\{|\mu_{k} - \xi_{k,\text{truth}}| \right\}.
\label{eqn:MAE}
\end{equation}

Next, we check for precision, which captures the amount of uncertainty on the predictions. We benchmark precision via the median size of the standard deviation, $\sigma_{k}$: 
\begin{equation}
    \text{M}(\sigma) = \text{median}_{k} \left\{\sigma_{k} \right\}.
\label{eqn:MS}
\end{equation}
Note that median error, MAE, and median $\sigma$ are evaluated separately for each parameter, as shown in Tables \ref{tab:metrics_shifted} and \ref{tab:metrics_doppel}.

After evaluating $\mu_{k}$ and $\sigma_{k}$ separately, we also need to check that the pair are calibrated properly. We assert that the ($\mu_{k}$,$\sigma_{k}$) predicted by the network describes Gaussian posteriors for the model parameters. To check whether this is true, we need to evaluate the calibration of the resulting posterior. For different $x$, we test whether $x\%$ of the posterior probability volume contains the ground truth $x\%$ of the time (see Figure \ref{calibration_curves}). The mathematical formalism for this metric is described in \citet{Park_2021} and \citet{Wagner_Carena_2021}. If $x\%$ of the probability volume contains the truth \textit{less} than $x\%$ of the time, the model posteriors are overconfident. If the opposite is true, the model posteriors are underconfident.

\begin{table*}
\begin{nolinenumbers}
\centering
\begin{tabular}{| c  c | c | c | c | c |}
    \hline
     & & \multicolumn{2}{c|}{\textbf{Individual Posteriors $p(\gamma_{\text{lens}}|d,\nu_{\text{int}})$}} & \multicolumn{2}{c|}{\textbf{Hyperposterior $p(\mathcal{M}_{\gamma_{\mathrm{lens}}}|\{d\})$}} \\
     \cline{3-6}
     & & $\gamma_{\text{lens}}$: \% error per lens & $\gamma_{\text{lens}}$: \% precision per lens & $\mathcal{M}_{\gamma_{\mathrm{lens}}}$: \% error & $\mathcal{M}_{\gamma_{\mathrm{lens}}}$: \% precision \\
        \hline
    \textbf{Shifted} & NPE & 4.6 & 6.1 & 2.9 & 1.9 \\
    & SNPE & 4.2 & 4.9 & 2.0 & 1.4\\
    \hline
    \textbf{Doppelganger} & NPE & 6.7 & 7.2 & 6.8 & 3.6 \\

    & SNPE & 5.0 & 5.4 & 4.4 & 2.3 \\
    \hline
\end{tabular}
\caption{Recovery of $\gamma_{\text{lens}}$ and $\mathcal{M}_{\gamma_{\text{lens}}}$ in verification tests. We highlight these results since error on $\gamma_{\text{lens}}$ directly translates to error on $H_0$ during TDC. We show results for the shifted test set (top row) and the doppelganger test set (bottom row). We evaluate performance at the individual lens modeling stage ($\gamma_{\text{lens}}$, assuming the interim prior) and the subsequent population-level stage ($\mathcal{M}_{\gamma_{\text{lens}}}$, assuming and inferring the conditional prior PDF). We also show how results compare for NPE and SNPE lens modeling. We find that the percent error decreases for both $\gamma_{\text{lens}}$ and $\mathcal{M}_{\gamma_{\text{lens}}}$ in both verification tests when we switch from NPE to SNPE modeling.} 
\label{tab:gamma_lens_metrics}
\end{nolinenumbers}
\end{table*}

Finally, we pay special attention to our performance on $\gamma_{\text{lens}}$. We compute the average percent error per lens and the average percent precision per lens. This is a useful benchmark, since percent error on $\gamma_{\text{lens}}$ translates to percent error on $H_0$ in TDC \citep{suyu2012cosmography}.

\subsection{Population Inference Metrics}
\label{subsection:hbi_metrics}

After analyzing each lens separately, individual mass models are combined to infer population-level properties. For each test set, individual posteriors are folded into a hierarchical inference for the hyperposterior $p(\nu | \{d\})$. 

The output of inference for $p(\nu | \{d\})$ is a set of 5e3 samples for each hyperparameter in $\nu$. We report the median of the samples along with an uncertainty derived from the averaged distance to the lower and upper 1$\sigma$ quantiles.
We compare the final values of $\nu$ to the ground truth, and report the error in units of standard deviation for all parameters in Tables \ref{shifted_hyperparams_table} and \ref{doppel_hyperparams_table}. For the doppelganger test set, we do not know the true population distribution the lens models are generated from. For these lenses, we take the sample mean and standard deviation, and use those values as a proxy ground truth we aim to recover. 

For this modeling stage, we also pay close attention to performance on $\gamma_{\text{lens}}$. In the hierarchical inference, our primary parameter of interest is the population mean: $\mathcal{M}_{\gamma_{\text{lens}}}$. We evaluate percent error and percent precision on this parameter in Table \ref{tab:gamma_lens_metrics}.

\subsection{Verification Tests}
\label{subsection:verification_results}

Before application to real data, we verify the performance of our method on two simulated test sets: the shifted set and the doppelganger set. See Section \ref{section:data} for a full description of these test sets. We compare performance using NPE and SNPE. 

\subsubsection{Shifted Test Set}

We start our verification process with the shifted test set. This set of 20 lenses is designed to test our ability to handle shifts between $\nu_{\text{int}}$ and the test distribution. First, we evaluate individual lens models. Parameter recovery of these models is summarized in Table \ref{tab:metrics_shifted}. The calibration of uncertainties is shown in Figure \ref{shifted_calib}. The calibration of both NPE and SNPE posteriors scatters about the perfect calibration curve within the 1$\sigma$ uncertainty region, indicating proper recovery of Gaussian $\mu_{k}$ and $\sigma_{k}$.

Next, we evaluate recovery of the population-level model. A portion of the posterior $p(\nu|\{d\})$ is shown in Figure \ref{shifted_contours}, with all inferred model parameters summarized in Table \ref{shifted_hyperparams_table}. We see that the error is less than 2$\sigma$ on all hyperparameters.

\subsubsection{Doppelganger Test Set}

After evaluating performance on the shifted set, we move to the doppelganger set. This test set consists of close mocks of the real HST lenses, and aims to test our ability to model realistic lensing configurations. Parameter recovery is summarized in Table \ref{tab:metrics_doppel}. We note that this performance is slightly worse than the shifted test set, which we will discuss in Section \ref{section:shifted_vs_doppels}. The calibration of Gaussian posteriors is summarized by the curves in Figure \ref{doppel_calib}. Both NPE and SNPE posteriors are overconfident, as evidenced by the departure from the perfect calibration line. This means that the predicted uncertainties $\hat{\sigma}_{k}$ are too small given the distances between $\hat{\mu}_{k}$ and $\xi_{\text{k,truth}}$. We discuss possible drivers of this effect in Section \ref{section:shifted_vs_doppels}. We continue on to the hierarchical modeling step.

\begin{figure*}[hbt!]
\begin{nolinenumbers}
\begin{center}
\subfloat[Shifted test set]{\includegraphics[scale=0.27]{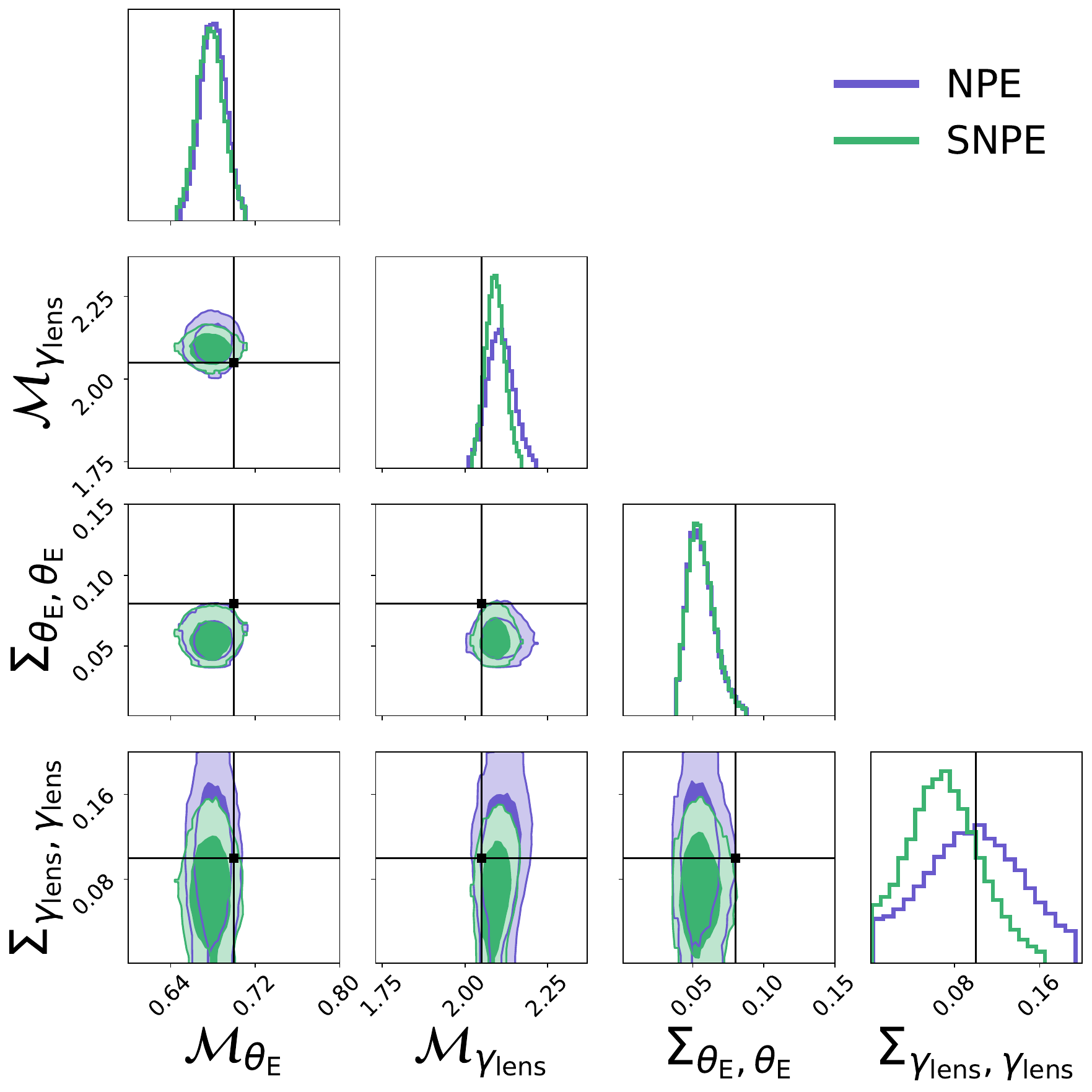}\label{shifted_contours}}
\quad
\subfloat[Doppelganger test set]{\includegraphics[scale=0.27]{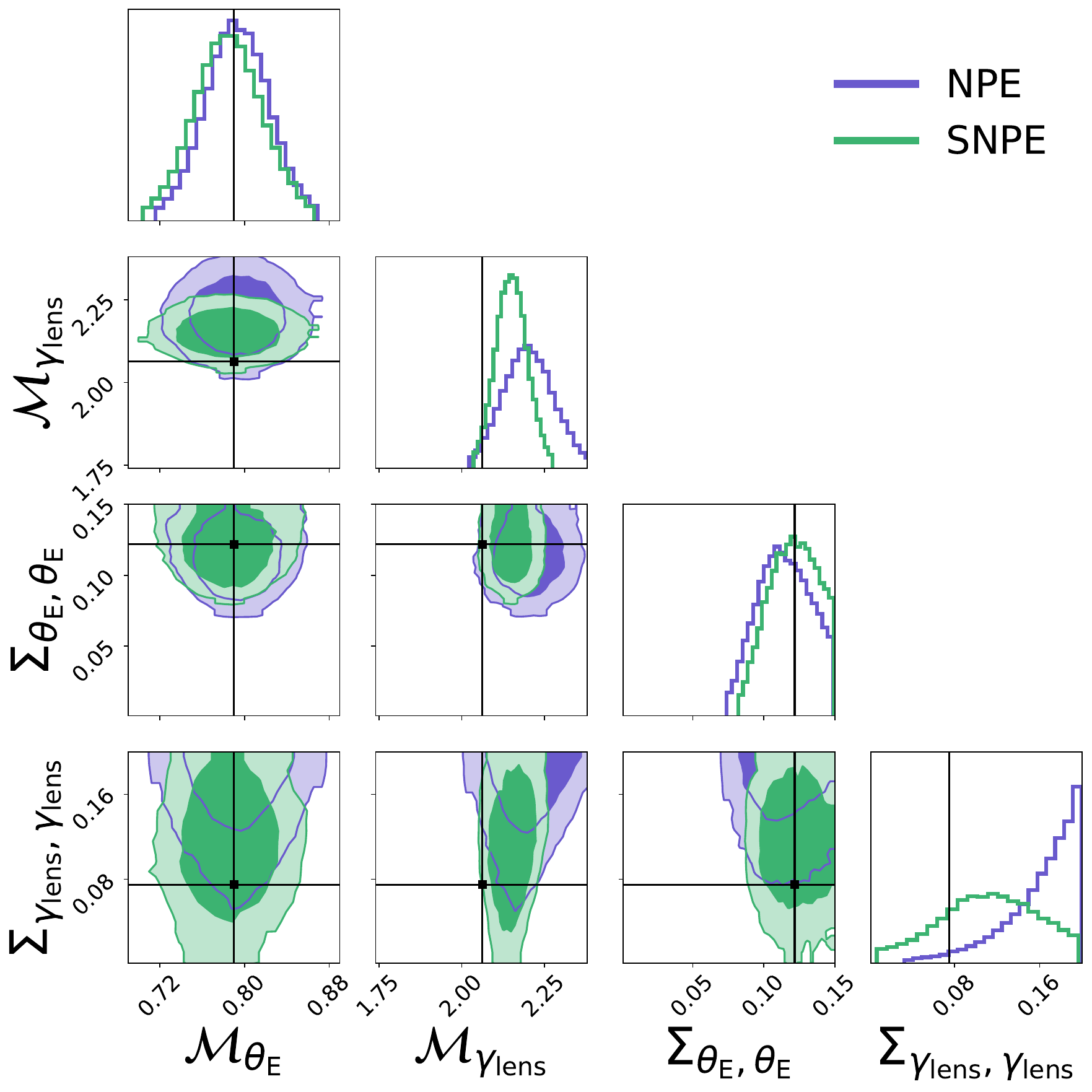}\label{doppel_contours}}
\caption{Two-dimensional contours of the hyperposterior p($\nu|\{d\}$) on the verification test sets. Inference from NPE posteriors is shown in purple. Inference from SNPE posteriors is shown in green. The ground truth is shown as a black line. Shaded contours are 68\% and 95\% intervals. Note for the doppelganger test set, we use the sample mean and standard deviation as a proxy ground truth.}
\end{center}
\end{nolinenumbers}
\end{figure*}

The hyperposterior is shown in Figure \ref{doppel_contours}, and recovery of hyperparameters is summarized in Table \ref{doppel_hyperparams_table}. Overconfident calibration of error bars means that during the hierarchical inference, uncertainty that should be attributed to individual posteriors is interpreted as a larger population $\sigma$.

Results from this doppelganger test set give us insight into possible systematic bias on $\gamma_{\text{lens}}$. We note that percent error on $\mathcal{M}_{\gamma_{\mathrm{lens}}}$ is higher than the percent precision for both shifted and doppelganger test sets. Performance on these sets is the best indicator of what we can expect on real data. When interpreting $\mathcal{M}_{\gamma_{\mathrm{lens}}}$ from the real data, we should keep in mind that the result might be affected by a small bias towards higher values.

\subsection{Application to HST Data}
\label{subsection:data_results}
After evaluating performance on verification tests, we apply our modeling technique to HST observations of 14 lensed quasars from the STRIDES sample. See Figure \ref{fig:hst_im} and Section \ref{section:hst_set} for more information on these lenses.

\begin{table*}[hbt]
\begin{nolinenumbers}
\centering
\begin{tabular}{| c | c | c | c | c | c | c |}
        \hline
        & $\mathcal{M}_{\theta_{\mathrm{E}}}$ & $\mathcal{M}_{\gamma_{\mathrm{lens}}}$ & $\Sigma_{\theta_{\mathrm{E}},\theta_{\mathrm{E}}}$ & $\Sigma_{\gamma_{1/2},\gamma_{1/2}}$ & $\Sigma_{\gamma_{\mathrm{lens}},\gamma_{\mathrm{lens}}}$ & $\Sigma_{e_{1/2},e_{1/2}}$\\
    \hline
    \textbf{Inferred} & 0.68 $\pm$ 0.01 & 2.09 $\pm$ 0.03 & 0.06 $\pm$ 0.01 & 0.12 $\pm$ 0.004 & 0.07 $\pm$ 0.04 & 0.16 $\pm$ 0.02\\
    \hline
    \;\;\;\; \textbf{Ground Truth} \;\;\;\; & 0.7 & 2.05 & 0.08 & 0.12 & 0.1 & 0.2\\
    \hline
    \textbf{Error (in $\sigma$)} & -2.0 & +1.3 & -2.0 & 0.0 & -0.8 & -2.0 \\
    \hline

\end{tabular}
\caption{Inferred hyperparameters $\nu$ for the \textbf{shifted set} using SNPE modeling. Error is reported in units of $\sigma$ to check for statistical significance.}
\label{shifted_hyperparams_table}
\end{nolinenumbers}
\end{table*}

\begin{table*}[hbt]
\begin{nolinenumbers}
\centering
\begin{tabular}{| c | c | c | c | c | c | c |}
        \hline
        & $\mathcal{M}_{\theta_{\mathrm{E}}}$ & $\mathcal{M}_{\gamma_{\mathrm{lens}}}$ & $\Sigma_{\theta_{\mathrm{E}},\theta_{\mathrm{E}}}$ & $\Sigma_{\gamma_{1/2},\gamma_{1/2}}$ & $\Sigma_{\gamma_{\mathrm{lens}},\gamma_{\mathrm{lens}}}$ & $\Sigma_{e_{1/2},e_{1/2}}$\\
    \hline
    \textbf{Inferred} & 0.78 $\pm$ 0.03 & 2.15 $\pm$ 0.05 & 0.12 $\pm$ 0.02 & 0.10 $\pm$ 0.01 & 0.11 $\pm$ 0.05 & 0.15 $\pm$ 0.02\\
    \hline
    \textbf{Proxy Ground Truth} & 0.79 & 2.06 & 0.12 & 0.08 & 0.08 & 0.12\\
    \hline
    \textbf{Error (in $\sigma$)} & -0.3 & +1.8 & 0.0 & +2.0 & +0.6 & +1.5 \\
    \hline
\end{tabular}
\caption{Inferred hyperparameters $\nu$ for the \textbf{doppelganger set} using SNPE modeling. Error is reported in units of $\sigma$ to check for statistical significance.}
\label{doppel_hyperparams_table}
\end{nolinenumbers}
\end{table*}

\begin{figure*}[hbt!]
    \centering
    \includegraphics[scale=0.65]{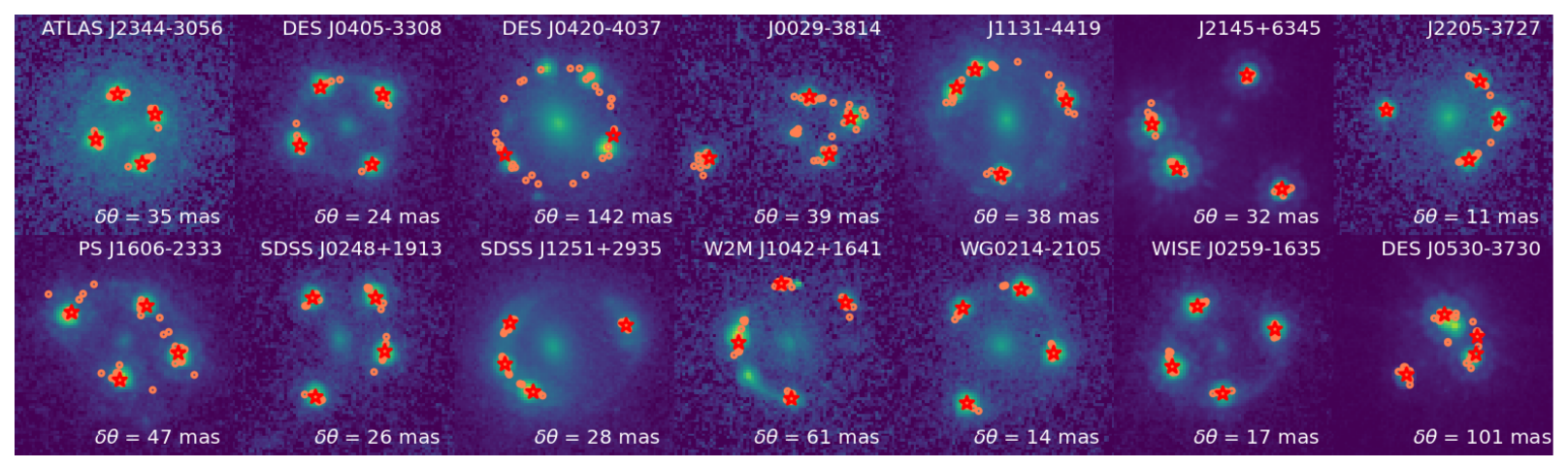}
    \caption{Image positions predicted by the SNPE lens models overlaid on top of the \textbf{real HST data}. HST images are 80$\times$80 pixels with 0.04$\arcsec$ resolution. Images are oriented with east to the left, and north to the top. All images are plotted with log-scaled color. Image positions computed from the mean of the lens model posterior are shown as red stars; image positions computed from 10 samples from the lens model posterior are shown as orange dots. Image positions computed from the mean of the posterior are compared against image positions from STRIDES23 modeling, and the average difference is quoted as $\delta\theta$ in milliarcseconds at the bottom corner of each image. If the SNPE-predicted image positions are farther than 0.14$\arcsec$ in either R.A. or decl. from any STRIDES23 image position, that image is discarded from the $\delta\theta$ calculation.}
    \label{fig:image_positions}
\end{figure*}

We perform both NPE and SNPE modeling of the data to produce approximate lens model posteriors. Since we have no ground truth, we are unable to verify NPE and SNPE performance using the metrics previously discussed.  Instead, we verify the modeling performance by calculating the image positions predicted by the SNPE lens model. Then, we overlay those image positions on top of the input data. This gives us an opportunity to roughly assess whether the lens model is consistent with the data. See Figure \ref{fig:image_positions}. We discuss recovery of point-source image positions in Section \ref{subsection:image_positions}.

We then perform a hierarchical inference for the population model using the 14 individual lens models. In Figure \ref{data_contours}, we show the population-level model from both NPE and SNPE posteriors. The hyperparameter values are reported in Table \ref{data_hyperparams_table}
for the inference using SNPE models. We use these values as our final result, since SNPE modeling had better performance on our primary metric of interest, $\mathcal{M}_{\gamma_{\text{lens}}}$, in verification tests. We find $\mathcal{M}_{\gamma_{\text{lens}}}=2.13  \pm 0.06$ on the real data, which we discuss further in Section \ref{discussion:mu_gamma_lens}.

\section{Discussion}
\label{section:discussion}
First, we discuss verification test results. We compare NPE and SNPE performance in Section \ref{subsection:NPE_vs_SNPE} and we discuss discrepancies between shifted set and doppelganger set performance in Section \ref{section:shifted_vs_doppels}. Then, we discuss results on real HST data, including image position recovery in Section \ref{subsection:image_positions}, context for our result on $\mu(\gamma_{\text{lens}})$ in Section \ref{discussion:mu_gamma_lens}, and a comparison to \citetalias{STRIDES} modeling in Section \ref{discussion:comp_to_FM}. After understanding results in more detail, we hypothesize possible  performance drivers such as the choice of functional form of the individual models in Section \ref{section:individ_model_functional_form} and the choice of functional form for the cPDF in Section \ref{section:cPDF_functional_form}. We elaborate on our attempt to use hierarchical re-weighting in Section \ref{section:reweighting_discussion}, and pose future directions in Section \ref{section:timing}. Finally, we report a timing analysis in Section \ref{discussion:future_directions}. 

\subsection{NPE vs SNPE}
\label{subsection:NPE_vs_SNPE}

We apply SNPE in an attempt to improve upon the precision of NPE posteriors. In an application of NPE to real data, we do not know the parameter range of the test distribution ahead of time, and thus must choose our training prior $\nu_{\text{int}}$ to be quite broad in order to fully cover the possible parameter space (see Table \ref{tab:training_prior}). We investigate SNPE as a way to cope with this large volume. We especially note that our training prior for the source's  $R_*$ is much wider than the one chosen in \cite{Park_2021}. Since the size of the source galaxy is known to have degeneracy with $\gamma_{\text{lens}}$ \citep{marshall_2007}, enforcing a wider prior for this parameter decreases our ability to break this degeneracy, and learn $\gamma_{\text{lens}}$.

We find some evidence for the success of SNPE when assessing performance on $\gamma_{\text{lens}}$ in Table \ref{tab:gamma_lens_metrics}. On both the shifted test set and the doppelganger test set, average percent error on $\gamma_{\text{lens}}$ decreased when switching from NPE to SNPE. However, if we look at the median error on $\gamma_{\text{lens}}$ for the shifted test set in Table \ref{tab:metrics_shifted}, a positive bias increases from 0.01 to 0.03 when switching from NPE to SNPE modeling. Here we caution that SNPE may amplify an underlying bias, but we cannot say for certain without further investigation into this effect.

\subsection{Shifted vs Doppelganger Performance}
\label{section:shifted_vs_doppels}

The only difference between the shifted test set and the doppelganger test set is the way in which the underlying model parameters were sampled. Shifted set lens parameters were sampled from a Gaussian distribution described in Appendix \ref{appendix:verification_tests}, while doppelganger set parameters come from forward modeling of real systems. We see some discrepancy between the performance on the two sets, most notably the overconfidence in error bars for the doppelganger set seen in Figure \ref{calibration_curves}. We have three hypotheses for this discrepancy. The first, and simplest, is that the doppelgangers are simply an unlucky statistical draw, since we are dealing with a relatively small sample size. The second hypothesis is that there is a complex selection function on the doppelgangers, putting them in a rarer part of parameter space compared to the shifted set. This selection may be difficult to see when interpreting in lower-dimensional space. The third is that the realistic lensing configurations of the doppelgangers are more sensitive to the diagonal covariance assumption. We test the use of full covariance posteriors in Appendix \ref{appendix:full_covariance}, and do see some evidence of better calibration on the doppelganger test set. See Section \ref{section:individ_model_functional_form} for a further discussion of full covariance NPE.

\begin{figure*}[hbt!]
    \centering
    \includegraphics[scale=0.35]{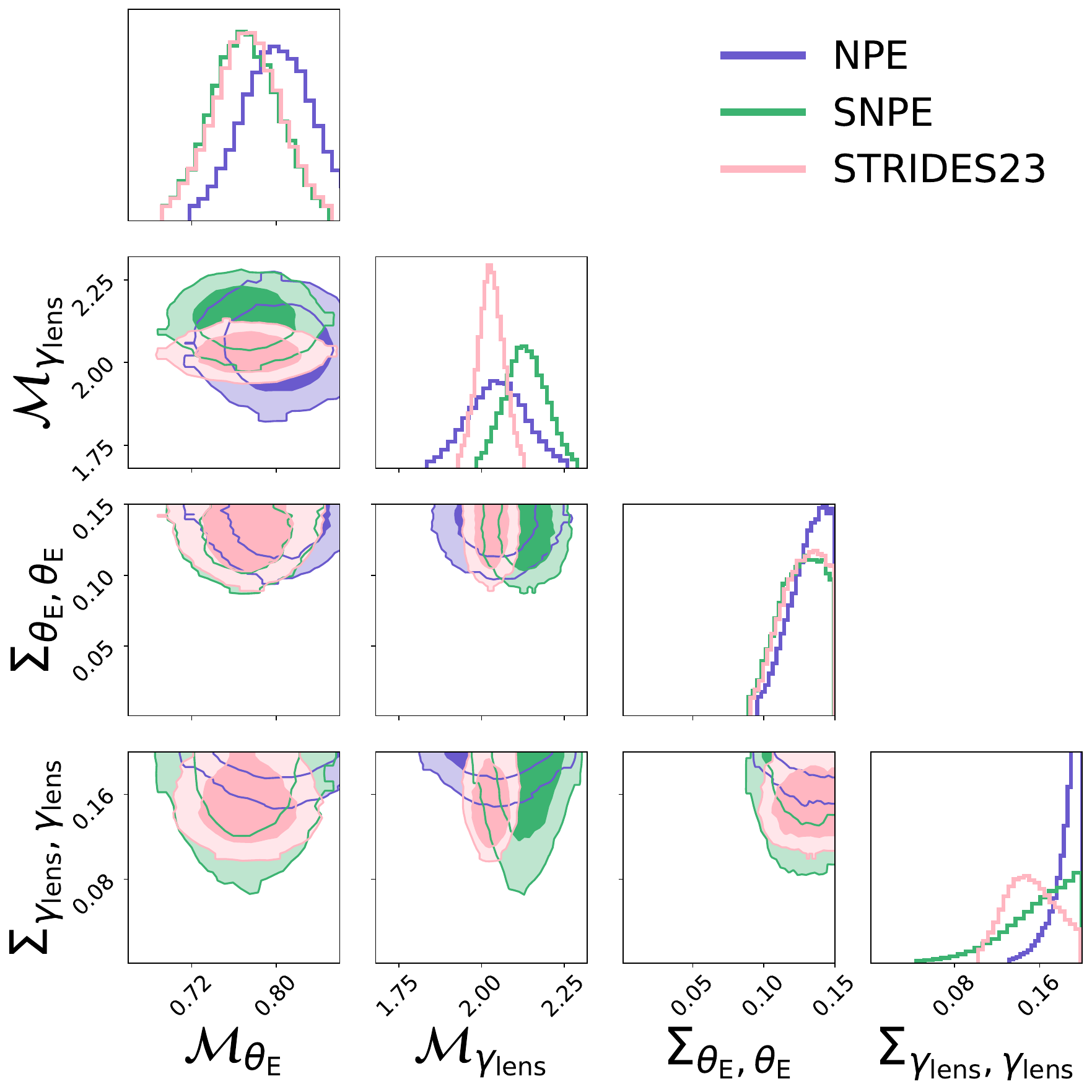}
    \caption{Two-dimensional contours of the hyperposterior p($\nu|\{d\}$) on the \textbf{real HST data}. Inference from NPE mass models is shown in purple. Inference from SNPE mass models is shown in green. Inference from STRIDES23 mass models is shown in pink. Shaded contours are 68\% and 95\% intervals. The NPE constraint on $\mu(\gamma_{\text{lens}})$ shifts and becomes more precise after applying SNPE, which is accompanied by a more constrained $\sigma(\gamma_{\text{lens}})$.}
    \label{data_contours}
\end{figure*}

\begin{table*}
\begin{nolinenumbers}
\begin{center}
\begin{tabular}{| c | c | c | c | c | c | c |}
    \hline
     & $\mathcal{M}_{\theta_{\mathrm{E}}}$ & $\mathcal{M}_{\gamma_{\mathrm{lens}}}$ & $\Sigma_{\theta_{\mathrm{E}},\theta_{\mathrm{E}}}$ & $\Sigma_{\gamma_{1/2},\gamma_{1/2}}$ & $\Sigma_{\gamma_{\mathrm{lens}},\gamma_{\mathrm{lens}}}$ & $\Sigma_{e_{1/2},e_{1/2}}$\\
    \hline
    \textbf{NPE} & 0.80 $\pm$ 0.03 & 2.05 $\pm$ 0.09 & 0.13 $\pm$ 0.01 & 0.11 $\pm$ 0.009 & 0.19 $\pm$ 0.01 & 0.18 $\pm$ 0.02 \\
    \hline
    \textbf{SNPE} & 0.77 $\pm$ 0.03 & 2.13 $\pm$ 0.06 & 0.13 $\pm$ 0.02 & 0.10 $\pm$ 0.01 & 0.16 $\pm$ 0.03 & 0.18 $\pm$ 0.02 \\
    \hline
    \;\;\;\;\;\; \textbf{STRIDES23} \;\;\;\;\;\; & 0.77 $\pm$ 0.03 & 2.03 $\pm$ 0.04 & 0.13 $\pm$ 0.02 & 0.08 $\pm$ 0.01 & 0.15 $\pm$ 0.03 & 0.14 $\pm$ 0.02 \\
    \hline
\end{tabular}
\caption{Inferred hyperparameters of the lens population model for the \textbf{real HST data}.}
\label{data_hyperparams_table}
\end{center}
\end{nolinenumbers}
\end{table*}

If any of these hypotheses are also true for the HST test set, our error bars on the data may be overconfidently calibrated. Overconfident mis-calibration of error bars has been found in forward modeling \citep{tan2023project}. We caution that mis-calibration of errors in individual posteriors can be absorbed by the population scatter when doing hierarchical inference, so careful error calibration is necessary to provide trustworthy population widths. In this work, we focus on the recovery of the population mean $\mathcal{M}_{\gamma_{\text{lens}}}$, and acknowledge that better error calibration will be needed in further work for $\Sigma_{\gamma_{\text{lens}},\gamma_{\text{lens}}}$.

\begin{figure*}
\centering
\includegraphics[scale=0.5]{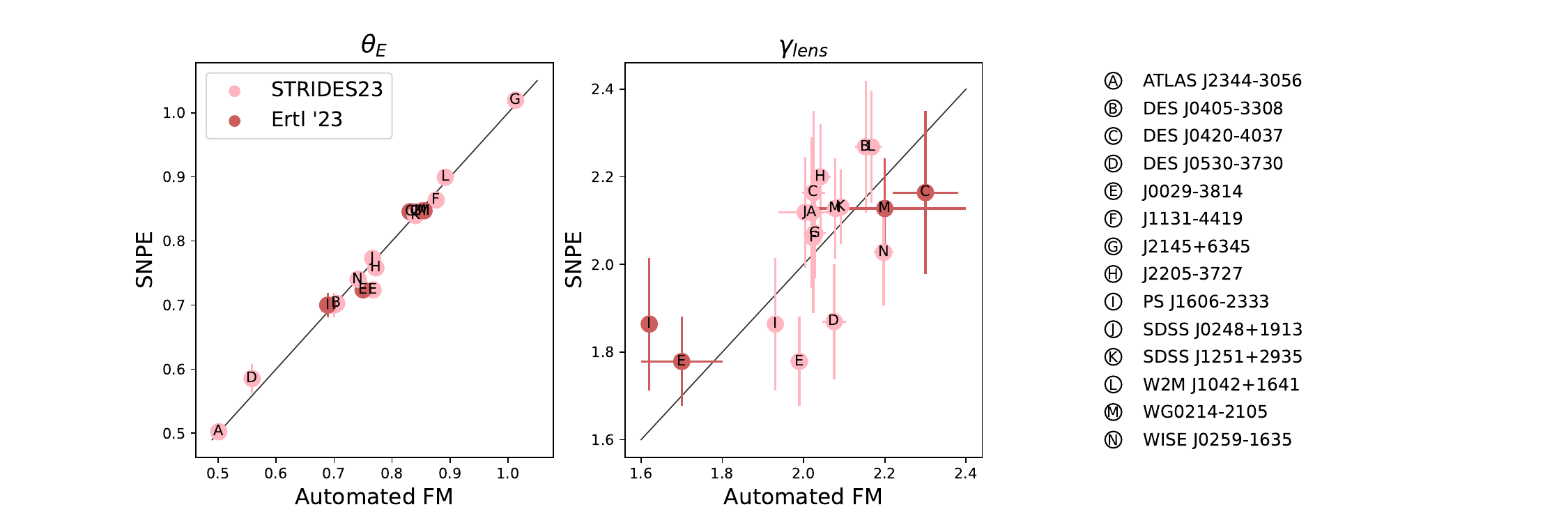}
\caption{Comparison of predicted mass model parameters $\theta_E$ and $\gamma_{\text{lens}}$ on \textbf{real HST data}. Comparison to \citetalias{STRIDES} models is shown in light pink; comparison to \citet{ertl2023tdcosmo} models is shown in red. The 1$\sigma$ uncertainty bars are shown for both NPE and automated forward modeling (FM) techniques. Note different priors and different model complexity are assumed for the separate modeling techniques. For comparisons of ellipticity and shear parameters, see Appendix \ref{appendix:ellip_shear}.}
\label{fig:comp_FM_params}
\end{figure*}

\subsection{Image Position Recovery on Data}
\label{subsection:image_positions}

We provide a benchmark for lens modeling performance on the real HST data that does not depend on ground truth lens model parameters. NPE lens models provide lens mass parameters and a source position, which is enough information to compute image positions. We compute the image positions from the predicted lens models, and overlay the positions on top of the input data. We compare the SNPE-predicted image positions to the image positions from the \citetalias{STRIDES} modeling (which we take as a proxy ground truth), and compute the average deviation $\delta\theta$. If the SNPE-predicted image positions are farther than 0.14$\arcsec$ in either R.A. or decl. from any \citetalias{STRIDES} image position, that image is discarded from the $\delta\theta$ calculation. We show this check in Figure \ref{fig:image_positions}. We discuss some possible causes of high $\delta\theta$. 

For lens DES J0420$-$4017, we hypothesize low performance on this lens is due to the complex source galaxy structure clearly visible in the lensed arc. The simulated training set did not include complex source light, so this problem could be mitigated with more realistic training simulations in the future. For lens W2M J1042+1641, there is an artifact in the top of the image in the F815W filter. The network seems to confuse this artifact for a point-source image, and ignores the point-source image in the bottom-left corner. This artifact is not present in observations from other filters, so we expect this effect could be mitigated by modeling in multiple filters simultaneously. If we remain limited to a single-band modeling tool, we could possibly mitigate this effect by adding similar artifacts to the noise model of our simulator. For lens DES J0530$-$3730, the lensing configuration is quite compact. The model produced by \citetalias{STRIDES} is a swallowtail lens, which is a rare configuration where the caustic overlaps. It is possible that during training, the network never saw a configuration close enough to this one to model the lens correctly. It may be possible to mitigate this problem with larger training sets or more efficient sampling of the parameter space during training.

In general, we try to make the NPE tool as robust as possible, but it is challenging to anticipate every feature that may appear in the data, such as the artifact seen in W2M J1042+1641. With enough lenses in our sample, we hope to be robust to edge cases such as this in the population-level analysis. Despite some lenses having visibly incorrect image positions, we use all 14 lenses for our hierarchical constraint. To check for robustness, we also perform the population-level inference without DES J0420$-$4017, W2M J1042+1641, and DES J0530$-$3730. We find that inclusion or exclusion of these lenses does not significantly change the constraint. For example, without the three lenses, we infer $\mathcal{M}_{\theta_{\mathrm{E}}} = 0.78 \pm 0.04 $ and $\mathcal{M}_{\gamma_{\mathrm{lens}}}  = 2.14 \pm 0.07 $. Comparing to our initial analysis in Table \ref{data_hyperparams_table}, we see these constraints are consistent.

\subsection{Population Distribution of the Power-Law Slope}
\label{discussion:mu_gamma_lens}

We are interested in our recovery of $\mathcal{M}_{\gamma_{\text{lens}}}$, since $\gamma_{\text{lens}}$ is known to be degenerate with $H_0$. Additionally, $\mathcal{M}_{\gamma_{\text{lens}}}$ has interesting astrophysical implications. On the real HST data, we measure a mean power-law slope of $\mathcal{M}_{\gamma_{\text{lens}}}=2.13  \pm 0.06 $. From automated forward modeling, we infer $\mathcal{M}_{\gamma_{\text{lens}}}=2.03  \pm 0.04 $.

Both values for the mean power-law slope are consistent with an isothermal profile, and are also consistent with the mean power-law slope at the Einstein radius measured on the SLACS lenses \citep{Auger_2010, Shajib_2021,etherington_automated_modeling,tan2023project}. This indicates a high degree of similarity between SLACS lenses and this subset of STRIDES deflectors, even though they have been selected differently and span over a different range of redshift. The fact that our measurement is compatible with an isothermal profile can be interpreted in the context of the ``bulge-halo" conspiracy \citep{treu_2006,Buote_2011,Cappellari_2016}, which refers to the nearly isothermal profiles of the total matter distribution observed in lensing and kinematic studies, although neither the stellar nor dark matter distribution follows a power law with $\gamma =2$. A specific arrangement between the two profiles is required to produce a profile that is nearly isothermal over a wide range of radius. We observe that a similar effect could be at play for this sample of lensed quasars.

\begin{figure*}[hbt!]
    \centering
    \includegraphics[scale=0.4]{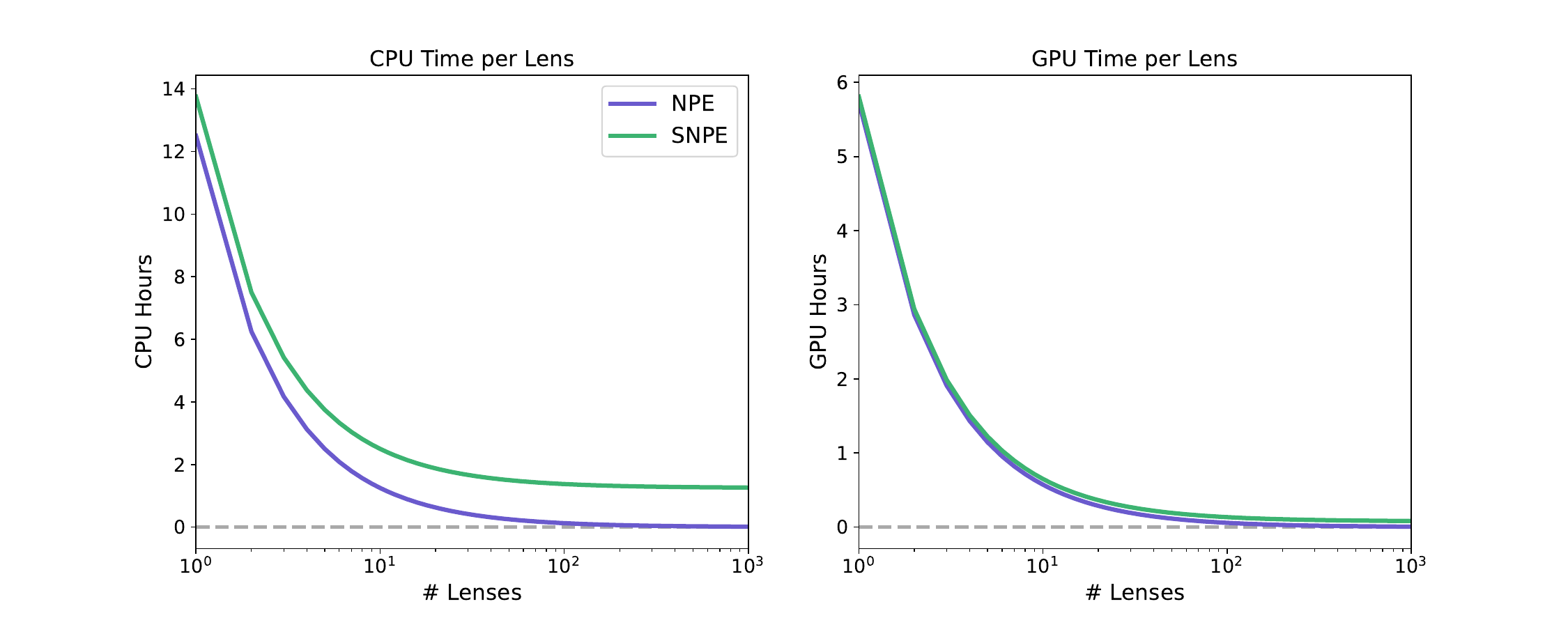}
    \caption{Lens modeling compute time as a function of test set size. We show the CPU and GPU compute time per lens. Note that the curves flatten as the number of lenses increases, since NPE training is an amortized cost. We assume 5e5 lenses are simulated for the NPE training set, and 5e4 lenses are simulated for each SNPE training set. Note that CPU compute time is bounded at 100 CPU hours per lens in STRIDES23. }
    \label{fig:timing_curves}
\end{figure*}

\subsection{Comparison to Automated Forward Modeling}
\label{discussion:comp_to_FM}

All of the HST lenses we model were also modeled in \citetalias{STRIDES} using an automated likelihood-based technique. Four lenses were additionally modelled in \citet{ertl2023tdcosmo}, also using an automated likelihood-based technique. We compare our results to these automated techniques on the data, but do so with an abundance of caution. There are some significant differences between the techniques, including: 

\begin{itemize}

\item \textbf{Model complexity}. In the automated forward modeling, more complex models are allowed. For example, in \citetalias{STRIDES}, lens light is assumed to have two components. In our NPE modeling, we assume single-component lens light. 

\item \textbf{Marginalization over nuisance parameters}. In automated forward modeling, every model component must be modelled exactly. For example, a choice of the PSF model must be made alongside the mass model, and all other model components. In our work, we do not require an explicit choice for the PSF (and other model components), but rather implicitly marginalize over the choice of PSF.

\item \textbf{Prior choices}. The modeling techniques assume different priors for parameters. This effect is most notable on $\gamma_{\text{lens}}$, which is often a prior-dominated parameter.

\end{itemize}

Despite these differences, we hope to find consistent posteriors between these modeling techniques. We compare the parameter estimates for $\theta_{\text{E}}$ and $\gamma_{\text{lens}}$ in Figure \ref{fig:comp_FM_params}. Additional comparisons for ellipticity and shear are included in Appendix \ref{appendix:ellip_shear}. We find that predictions for $\theta_{\text{E}}$ are consistent. On the other hand, predictions for $\gamma_{\text{lens}}$ are not always consistent. In particular, while the discrepancy between modeling methods is often captured by the larger NPE 1$\sigma$ uncertainty, the \citetalias{STRIDES} 1$\sigma$ uncertainty is not large enough to account for any scatter in modeling results. This is further evidence for the conclusion presented in \citet{tan2023project} that the uncertainty from forward modeling techniques is often underestimated. The larger $
\sigma$ values from \citet{ertl2023tdcosmo} may more accurately reflect uncertainty from the automated lens modeling. 

We also compare our population model to the population model inferred from \citetalias{STRIDES} posteriors, as shown in Figure \ref{data_contours} and Table \ref{data_hyperparams_table}. We explicitly account for the informative prior on $\gamma_{\text{lens}}$ that was assumed for the \citetalias{STRIDES} modeling by including an interim prior $\nu_{int}$ in our hierarchical inference with the forward modeling posteriors. The models agree perfectly for $\mathcal{M}_{\theta_{\mathrm{E}}}$ and $\Sigma_{\theta_{\mathrm{E}},\theta_{\mathrm{E}}}$. There is some discrepancy on $\mathcal{M}_{\gamma_{\text{lens}}}$, which is not surprising considering the scatter on individual predictions seen in Figure \ref{fig:comp_FM_params}. However, central values are consistent within 2$\sigma$. Additionally, the precision on $\mathcal{M}_{\gamma_{\text{lens}}}$ is comparable between the two techniques (3\% with SNPE modeling, and 2\% with automated forward modeling).

\subsection{Lens Model Functional Form}
\label{section:individ_model_functional_form}

Assuming a diagonal Gaussian functional form for the lens model posteriors is a simplification. We do expect covariances between some lens parameters, which our model cannot capture. We suspect that when there is not enough information in an image to break degeneracies between parameters, our diagonal posterior is forced to overconfidently converge to one solution rather than widen to allow both solutions. Ultimately, we suspect that the lack of expressivity of our functional form is one of the main weaknesses of this method.

We experimented with allowing a full covariance in our Gaussian posteriors, thereby enabling the lens models to capture correlations between parameters. This technique passed both verification tests, with similar performance to the diagonal NPE. However, when we applied the full covariance NPE to the 14 real HST lenses, we saw multiple predictions with a $\sigma_k(\gamma_{\text{lens}})$ values wider than the $\gamma_{\text{lens}}$ training distribution. We hypothesize that with the freedom to predict a full covariance matrix, the model is less robust to the domain shift between simulated and real data. We suggest that when increasing the flexibility of the functional form, an accompanying increase in the realistic complexity in the training simulations is necessary. \footnote{Note that since the SNPE proposal distribution depends on the NPE posterior, this behavior prevented us from running full covariance SNPE.} For more details on this test, see Appendix \ref{appendix:full_covariance}. We note that there is evidence that a more flexible functional form can improve performance. In one application of neural network modeling to galaxy-galaxy lenses, it was demonstrated that calibration of posteriors was improved when moving from a single Gaussian posterior to a mixture of Gaussians \citep{Legin_2023}.

\subsection{Population Model Functional Form}
\label{section:cPDF_functional_form}

For this analysis, we assumed the population distribution of lens parameters $p(\xi|\nu)$ takes a diagonal Gaussian functional form. The true population distribution almost certainly follows a more complex functional form. For example, we know the distribution of $\theta_E$ has a heavy truncation at 0, and its shape is non-Gaussian. In future analyses, we aim to relax the Gaussian assumption. 

\subsection{Re-Weighting Individual Posteriors}
\label{section:reweighting_discussion}

When using lens models within a hierarchical framework, the individual systems' posterior PDFs should be computed by re-weighting the interim posteriors, accounting for the distribution shift between the interim prior and the conditional PDF in the process. This hierarchical re-weighting scheme was derived in \cite{Wagner_Carena_2021}, and the idea is that each lens model can take advantage of additional information from the population-level model. We applied the same re-weighting scheme for our individual posteriors, with some modifications given the small number of lenses in our sample. For details, see Appendix \ref{appendix:reweighting}. 

We found that the resulting re-weighted, ``final'' posterior PDFs were systematically overconfident in our verification tests. This could stem from mis-calibration in the original posteriors that is amplified after re-weighting. Our hypothesis is that this is a likely the result of an insufficiently flexible functional form of the lens posteriors. We save the investigation of more flexible PDFs for further work.

\subsection{Timing}
\label{section:timing}

NPE and SNPE are advantageous modeling techniques given their speed compared to likelihood-based techniques. We analyze the computing resources required for our method, and project how this requirement will scale for the number of lenses in a given test set. We take stock of the compute used for each step of the modeling process. Then, we summarize the total compute used per lens. Simulation of lenses using \textsc{paltas} takes 0.09 CPU seconds per lens. Simulation cost is incurred one time for the NPE training set (12.5 hours for 5e5 lenses), and many times for SNPE training (1.25 hours for 5e4 lenses), which requires a new training set for every lens. Training of the network uses GPU compute. NPE training is only run once, and converges in 5.73 GPU hours. SNPE training is run once for each lens, and uses an additional 0.46 GPU hours per lens. Finally, we use CPU compute to generate predictions. This contribution is negligible, as it only takes 0.7 seconds per lens. We plot the compute time as a function of the number of lenses in the test set in Figure \ref{fig:timing_curves}. Note that if we increase the number of lenses in the test set, this decreases the cost of the initial NPE training per lens, but the SNPE cost remains the same.

\subsection{Future Directions}
\label{discussion:future_directions}

We would like to relax the assumption of a Gaussian functional form for the lens model posteriors. An extension to this method would be to use a normalizing flow component to allow freedom of the functional form of the lens posterior, as is done in \citet{poh2022strong}.

Instead of using SNPE, we could use the NPE approximate posterior as a starting point for likelihood-based forward modeling. This approach was examined in \citet{Pearson_2021}.

The primary driver of performance with machine learning techniques is often the quality of the training data. To improve the quality of training examples, we should add more realism to our training set simulations. One important factor may be the assumed source galaxy light profile in our training simulation, which is a single S\'ersic. More complexity could be included in future applications of this technique. For example, we could introduce postage stamps of real galaxies for the source galaxy light in training simulations. This is done using COSMOS galaxies in \citet{Wagner_Carena_2023}. Going futher, we could use postage stamps of real galaxies for both the source galaxy and lens light components, as is done in \citet{Schuldt_2023}. We could also increase the complexity of the mass model, for example, including massive satellites, which we know are present in lenses like PS J1606$-$2333.

\section{Conclusion}
\label{section:conclusion}

We are working to enable an independent measurement of the Universe's expansion through TDC with lensed quasars discovered by LSST. To enable population-level analyses of a large sample of time-delay lenses, we apply NPE and SNPE for fast and standardizable lens modeling. We use verification tests to establish confidence in our approximate inference technique, ensuring both individual lens models and the population-level model are recovered. After assessing verification tests, we apply our technique to real time-delay lenses for the first time. We put the first population-level constraint on a subset of the STRIDES lensed quasars, finding $\mathcal{M}_{\gamma_{\text{lens}}}=2.13  \pm 0.06 $.

We address the guiding questions we initially posed:

\begin{itemize}

    \item Will the application of NPE for strong lens mass modeling produce reliable models on real time-delay lenses? In verification tests, what is the percent error per lens on the PEMD power-law slope $\gamma_{\text{lens}}$?

    Answer: NPE and SNPE modeling are successfully used for our first application on HST data. Individual lens constraints recovered during verification tests have roughly 5\% error per lens on $\gamma_{\text{lens}}$.

    \item How does the application of SNPE compare to NPE? Does the increased sampling density of SNPE improve the precision of lens model posteriors?

    Answer: SNPE improves percent error at the individual level and the population level for $\gamma_{\text{lens}}$. However, it is unclear that SNPE always improves performance, as evidenced by MAE.

    \item What population constraint can we put on the real data using HBI? In verification tests, what is the percent error on the population mean of the PEMD power-law slope $\mathcal{M}_{\gamma_{\text{lens}}}$?

    Answer: We achieve 2\% and 4\% errors on $\mathcal{M}_{\gamma_{\text{lens}}}$ in shifted and doppelganger verification tests, respectively, with SNPE modeling. We find a population mean on the real HST data: $\mathcal{M}_{\gamma_{\text{lens}}}=2.13  \pm 0.06 $.

\end{itemize}

With this work, we make necessary steps towards building a fully automated lens modeling pipeline that will be consistently applied to all parts of the LSST strong lens sample. 

\section{Acknowledgements}

This paper has undergone internal review in the LSST Dark Energy Science Collaboration. SE would like to thank internal reviewers Stefan Schuldt and Anowar Shajib for their extensive contributions in this role. We additionally thank Jelle Aalbers, Phil Holloway, Xiangyu Huang, Ralf Kaehler, Narayan Khadka, Tian Li, and Greg Madejski for many useful conversations along the way. We thank Ji Won Park, Tom Collett, Sherry Suyu, and Aymeric Galan for impactful conversations and suggestions. We thank the journal referee for providing valuable feedback. 

SE developed all code in \textsc{lens-npe}, ran all analysis, produced all figures, and wrote the main body of the text. SWC helped design and implement the method and provided feedback on all aspects. PM helped design the statistical framework and verification tests, and provided feedback on all aspects. MM provided interpretation and context for results and provided feedback on all aspects. SB helped with simulation implementation and provided feedback on all aspects. AR provided feedback on all aspects. TS and TT reduced HST data products, produced forward modeling chains, and provided help handling the data products, in addition to feedback on all results. SS and AJS provided feedback and suggestions on all aspects. PV provided feedback on method and results. 

HST observations were conducted by programs HST-GO-15320 and HST-GO-15652 (PI: Treu).
SE acknowledges funding from the National Science Foundation GRFP, and the Stanford Data Science Scholars program. 
This work was supported by the U.S. Department of Energy under contract No. DE-AC02-76SF00515.
MM acknowledges support by the SNSF (Swiss National Science Foundation) through mobility grant P500PT\_203114.
SB acknowledges support by the Department of Physics and Astronomy, Stony Brook University.
TS and TT acknowledge support by the the National Science Foundation through grant NSF-AST-1906976 and NSF-AST-1907396 ``Collaborative Research: Toward a 1\% measurement of the Hubble Constant with gravitational time-delays". 
SS has received funding from the European Union’s Horizon 2022 research and innovation programme under the Marie Skłodowska-Curie grant agreement No 101105167 - FASTIDIoUS.
Support for this work was provided by NASA through the NASA Hubble Fellowship grant HST-HF2-51492 awarded to A.J.S. by the Space Telescope Science Institute, which is operated by the Association of Universities for Research in Astronomy, Inc., for NASA, under contract NAS5-26555. A.J.S. also received support from NASA through STScI grants HST-GO-16773 and JWST-GO-2974.
This work was also supported by NASA through the NASA Hubble Fellowship grant HST-HF2-51492 awarded to A.S. by the Space Telescope Science Institute, which is operated by the Association of Universities for Research in Astronomy, Inc., for NASA, under contract NAS5-26555.

The DESC acknowledges ongoing support from the Institut National de 
Physique Nucl\'eaire et de Physique des Particules in France; the 
Science \& Technology Facilities Council in the United Kingdom; and the
Department of Energy and the LSST Discovery Alliance
in the United States.  DESC uses resources of the IN2P3 
Computing Center (CC-IN2P3--Lyon/Villeurbanne-France) funded by the 
Centre National de la Recherche Scientifique; the National Energy 
Research Scientific Computing Center, a DOE Office of Science User 
Facility supported by the Office of Science of the U.S.\ Department of
Energy under contract No.\ DE-AC02-05CH11231; STFC DiRAC HPC Facilities, 
funded by UK BEIS National E-infrastructure capital grants; and the UK 
particle physics grid, supported by the GridPP Collaboration.  This 
work was performed in part under DOE contract DE-AC02-76SF00515. 

Source code for this work is publicly available in the repository \textsc{lens-npe}\footnote{\url{https://github.com/smericks/lens-npe}} \citep{erickson_lens-npe}. This repository makes use of public software packages \textsc{paltas} \citep{Wagner_Carena_2023} and \textsc{lenstronomy} \citep{birrer2018lenstronomy, birrer2021lenstronomy}. Model weights, predictions, and chains used for Figures 5, 6, 8, 9, 11, 12, 13, 14, 15, 16 are provided in Zenodo at \dataset[doi: 10.5281/zenodo.13906030]{https://doi.org/10.5281/zenodo.13906030}.

\bibliography{main}

\appendix

\section{Training Prior}
\label{training_prior}

Our choice for the interim training prior is shown in Table \ref{tab:training_prior}. When training NPE for scientific application, $\nu_{\text{int}}$ must be chosen carefully, such that any possible lensing configuration is included in the prior volume. The assumed model for each component is discussed in Section \ref{section:modeling_choices}.

\begin{table*}[hbt!]
\begin{nolinenumbers}
\begin{center}
\begin{tabular}{| c | c | c |}

    \hline
     \textbf{Main Deflector} 
     & $\theta_E (\arcsec)$ & $\mathcal{N}_{\text{trunc}}(min=0,\mu=0.8,\sigma=0.15)$\\

     & $\gamma_{\text{lens}}$ & $\mathcal{N}(\mu=2.0,\sigma=0.2)$ \\

     & $e_1, e_2$ & $\mathcal{N}(\mu=0.0,\sigma=0.2)$\\

     & $x_{\text{lens}}, y_{\text{lens}} (\arcsec)$ & $\mathcal{N}(\mu=0,\sigma=0.07)$\\

     & $\gamma_1, \gamma_2$ & $\mathcal{N}(\mu=0,\sigma=0.12)$\\

    \hline
     \textbf{Lens Light} & $m_{\text{app}}$ & $\mathcal{N}_{\text{trunc}}(min=17,max=23,\mu=20,\sigma=2)$ \\

     & $R_* (\arcsec)$ & $\mathcal{N}_{\text{trunc}}(min=0,\mu=1,\sigma=0.8)$\\

     & $n_*$ & $\mathcal{N}_{\text{trunc}}(min=0.5,\mu=3,\sigma=2)$ \\

     & $e_1, e_2$ & $\mathcal{N}_{\text{trunc}}(min=-0.5,max=0.5,\mu=0.0,\sigma=0.2)$\\

     & $x_{\text{ll}} (\arcsec)$ & $\mathcal{N}(\mu=x_{\text{lens}},\sigma=0.005)$\\

     & $y_{\text{ll}} (\arcsec)$ & $\mathcal{N}(\mu=y_{\text{lens}},\sigma=0.005)$\\

    \hline

     \textbf{Source Light} & $m_{\text{app}}$ & $\mathcal{N}_{\text{trunc}}(min=20,max=27,\mu=23.5,\sigma=1.7)$ \\

     & $R_* (\arcsec)$ & $\mathcal{N}_{\text{trunc}}(min=0,\mu=0.5,\sigma=0.8)$\\

     & $n_*$ & $\mathcal{N}_{\text{trunc}}(min=0.5,\mu=3,\sigma=2)$ \\

     & $e_1, e_2$ & $\mathcal{N}_{\text{trunc}}(min=-0.5,max=0.5,\mu=0.0,\sigma=0.2)$\\

     & $x_{\text{src}} (\arcsec)$ & $\mathcal{N}(\mu=0,\sigma=0.1)$\\

     & $y_{\text{src}} (\arcsec)$ & $\mathcal{N}(\mu=0,\sigma=0.1)$\\

    \hline   
     \textbf{Point Source} & $m_{\text{app}}$ & $\mathcal{N}_{\text{trunc}}(min=19,max=25,\mu=22,\sigma=2)$ \\

     & $x_{\text{ps}} (\arcsec)$ & $x_{\text{src}}$\\

     & $y_{\text{ps}} (\arcsec)$ & $y_{\text{src}}$\\

     & $f_{\text{microlensing}}$ & $\mathcal{N}_{\text{trunc}}(min=0,\mu=1.0,\sigma=0.3)$\\
    \hline

\end{tabular}
\caption{Choice of the interim training prior, $\nu_{\text{int}}$. Note $\mathcal{N}$ is a Gaussian, and $\mathcal{N}_{\text{trunc}}$ is a truncated Gaussian. The main deflector is parameterized by a PEMD and external shear profile (see Section \ref{subsubsection:mass_profile}). Lens light and source light are parameterized by a S\'ersic profile (see Section \ref{subsubsection:light_profile}). We define the prior on lens light, source light, and point-source brightness in apparent magnitude, $m_{\text{app}}$. We apply a fractional change to each point-source image's apparent magnitude, $f_{\text{microlensing}}$, to account for microlensing.} 
\label{tab:training_prior}
\end{center}
\end{nolinenumbers}
\end{table*}

\begin{figure*}[hbt!]
    \begin{nolinenumbers}
    \centering
    \includegraphics[scale=0.5]{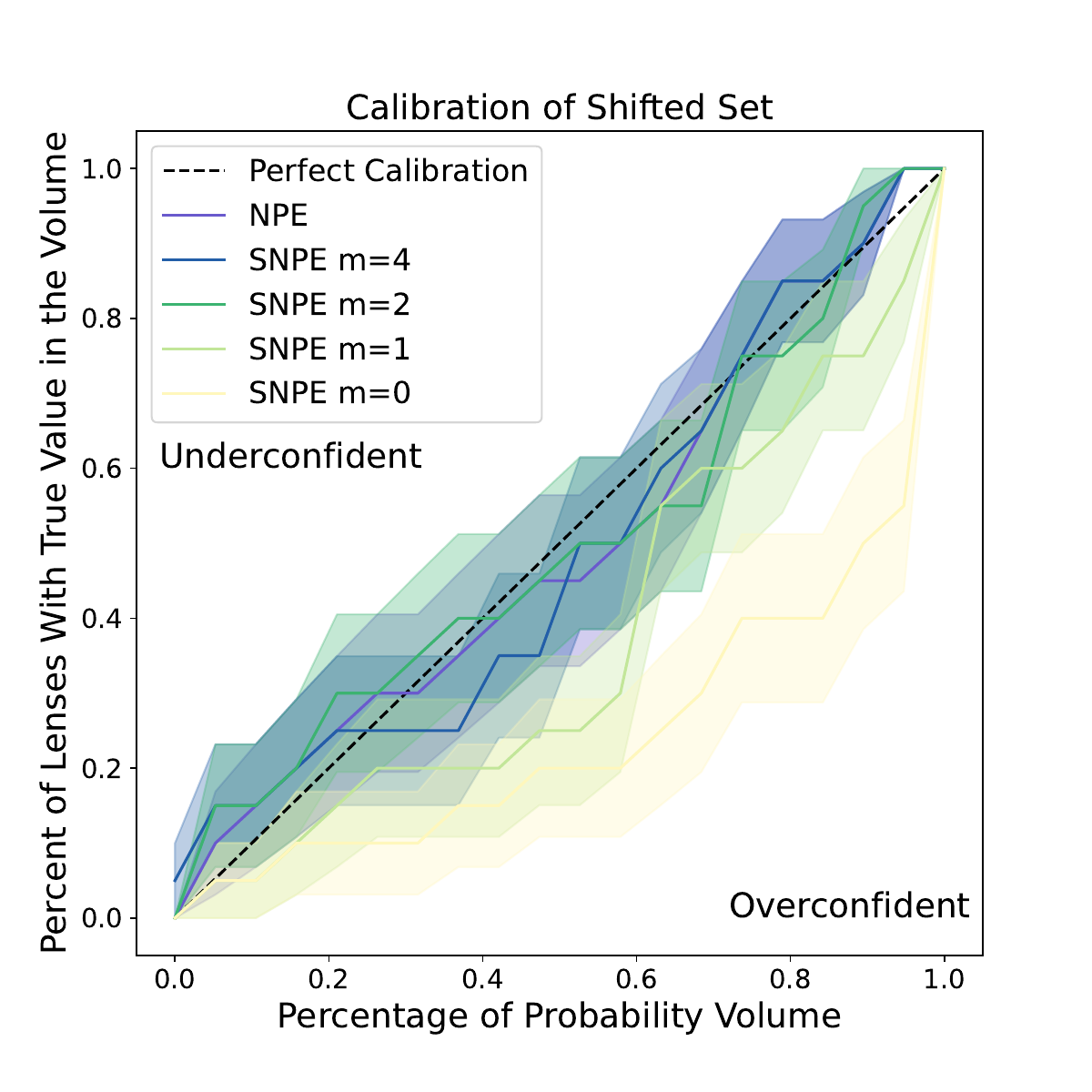}
    \caption{Calibration curves for the shifted test set. In perfectly calibrated posteriors (dashed line), a given x\% of the probability volume contains the truth x\% of the time. Calibration of NPE posteriors is shown in blue-purple. Calibration of SNPE posteriors with different geometrically averaged proposals is shown in blue (m=4), green (m=2), yellow-green (m=1), and yellow (m=0). Note that m=2 is the choice made for this analysis. The shaded region encompasses 1$\sigma$ uncertainty.}
    \label{gem_avg_calib}
    \end{nolinenumbers}
\end{figure*}

\section{Verification Tests}
We describe in more detail how we simulate the verification test sets. 

\label{appendix:verification_tests}
\subsection{Shifted Distribution}

To create the shifted test set, we need to define a distribution from which to sample lens parameters. We start with the training prior, detailed in Table \ref{tab:training_prior}. Then, we shift and narrow the distribution in two key parameters: $\theta_E$ and $\gamma_{\text{lens}}$. We choose the shifts to mimic the distribution shift we see in the doppelganger test set. We change the distribution for these two parameters as follows: 
\begin{equation}
    \theta_E \sim \mathcal{N}_{\text{trunc}}(min=0,\mu=0.7,\sigma=0.08)
\end{equation}
\begin{equation}
    \gamma_{\text{lens}} \sim \mathcal{N}(\mu=2.05,\sigma=0.1).
\end{equation}
All other parameters are sampled in the same way as the training set, and the same simulator is used. We draw 20 lenses, both double and quad configurations, from this distribution to make the narrow test set.

\subsection{Doppelganger Simulations}
\label{appendix:doppels}

To create the doppelganger test set, we aim to mimic the STRIDES dataset with simulations. The lensing parameters for these simulations are taken from the existing \citetalias{STRIDES} models. These models have more components than our simulations. We simplify the models to match the parameterization of our training simulations. For example, some \citetalias{STRIDES} models have double component lens light. To simplify these models, we use only the bulge component for our doppelganger simulations. Another key difference is that we input the source position ($x_{\text{src}}$,$y_{\text{src}}$) to our simulator, but the \citetalias{STRIDES} models do not provide a source position, but rather fit for image positions in the lens plane: ($x_{\text{img}}$,$y_{\text{img}}$). When making these doppelgangers, we take the image positions and mass model from \citetalias{STRIDES}, and solve the lens equation for the source position using the numerical solver in \textsc{lenstronomy}. Note that when simulating the doppelganger of DES J0530$-$3730 with this procedure, we were unable to find a single source position solution that matched all four point-source images, so we exclude this lens from our doppelganger test set. Note that the \citetalias{STRIDES} lens model for DES J0530$-$3730 has a rare swallowtail configuration, which is very sensitive to the source position.

\section{Geometric Averaging of SNPE Proposals}
\label{appendix:geom_averaging}

Given the training prior $p(\xi_{k})$ and the NPE approximate posterior $q_\phi(\xi_{k}|d_{k},\nu_{\text{int}})$, we need to choose a proposal distribution $\tilde{p}(\xi_{k})$ to generate new training examples for SNPE. We investigate a proposal that uses a geometric average of the prior and the NPE approximate posterior: 
\begin{equation}
    \tilde{p}(\xi_{k}) \propto \left( q_\phi(\xi_{k}|d_{k},\nu_{\text{int}})^n p(\xi_{k}|\nu_{\text{int}})^m \right)^{\frac{1}{m+n}}
\end{equation}

Parameter $n$ is the weight of the posterior in the average, and $m$ is the weight of the prior in the average. We keep $n$=1, but we allow $m$ to take values [0,1,2,4].

We test different values of $m$ on the shifted test set. We show the calibration of shifted set posteriors for different values of $m$ in Figure \ref{gem_avg_calib}. We choose $m$=2 for our main analysis, since at this value we strike a balance of folding in as much information from the NPE prediction as possible without causing overconfident calibration.

\begin{figure*}[hbt!]
    \begin{nolinenumbers}
    \centering
    \includegraphics[scale=0.5]{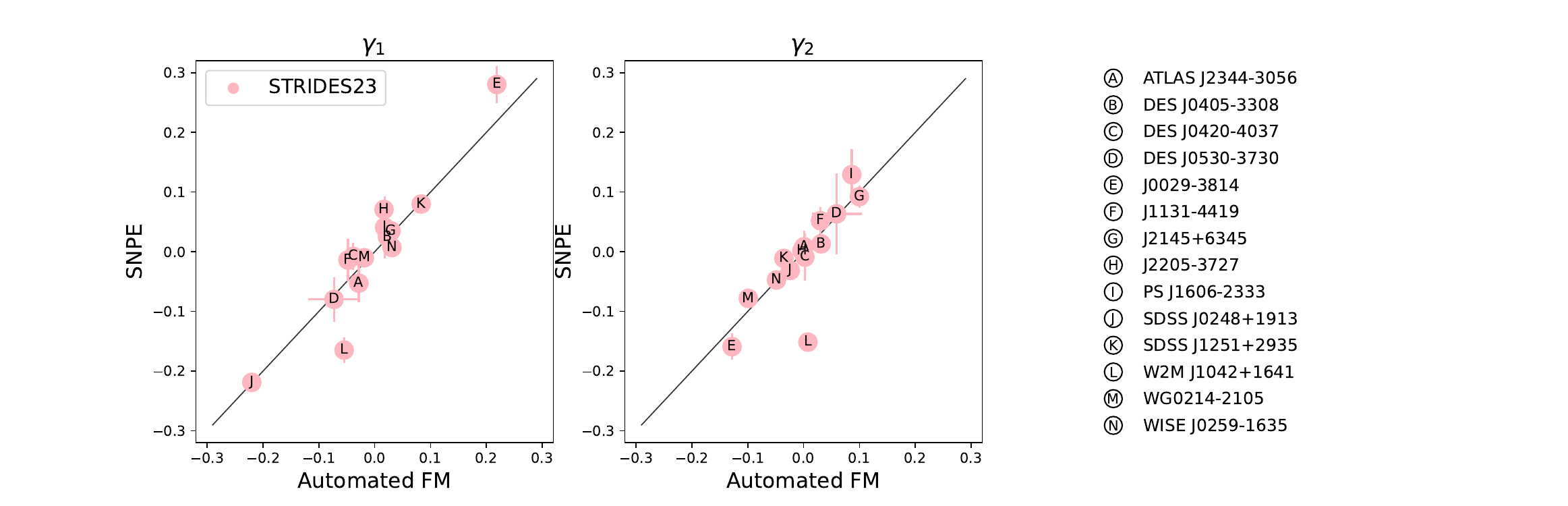}
    \caption{Comparison of predicted mass model parameters $\gamma_1$ and $\gamma_2$ on \textbf{real HST data}. Comparison to \citetalias{STRIDES} models is shown in light pink. 1$\sigma$ uncertainty bars are shown for both NPE and automated forward modeling (FM) techniques. Note different priors and different model complexity are assumed for the separate modeling techniques.}
    \label{external_shear_comp}
    \end{nolinenumbers}
\end{figure*}

\begin{figure*}[hbt!]
    \begin{nolinenumbers}
    \centering
    \includegraphics[scale=0.5]{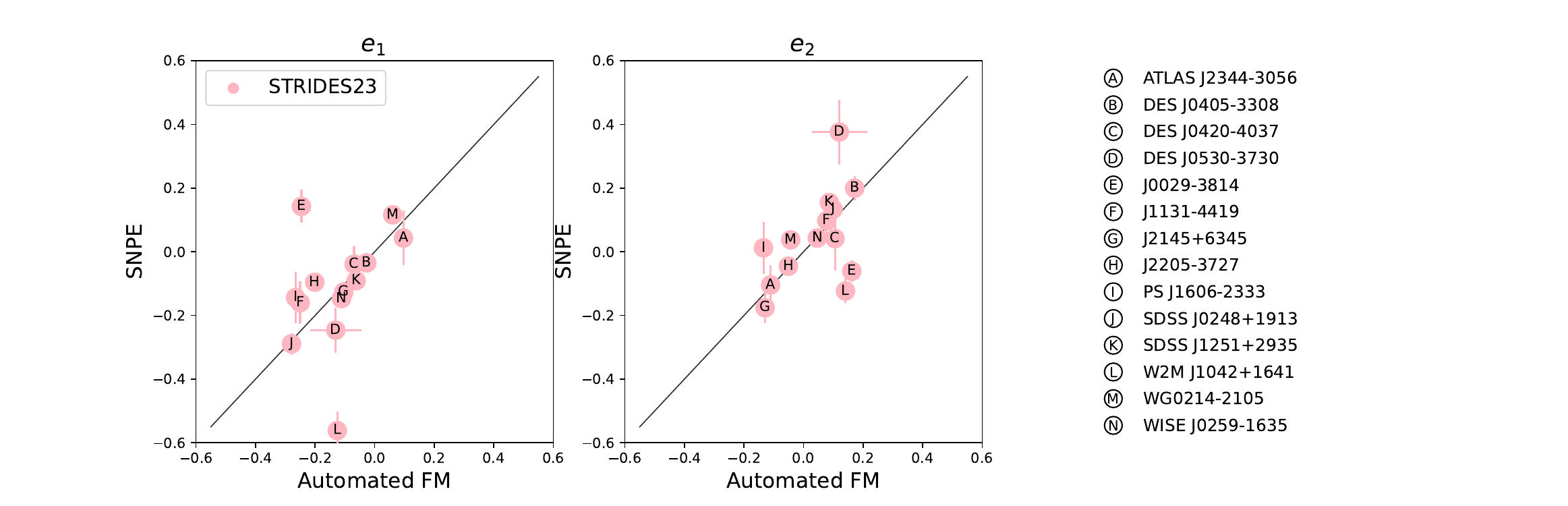}
    \caption{Comparison of predicted mass model parameters $e_1$ and $e_2$ on \textbf{real HST data}. Comparison to \citetalias{STRIDES} models is shown in light pink. 1$\sigma$ uncertainty bars are shown for both NPE and automated forward modeling (FM) techniques. Note different priors and different model complexity are assumed for the separate modeling techniques.}
    \label{ellipticity_comp}
    \end{nolinenumbers}
\end{figure*}

\section{Additional Lens Parameter Comparisons}
\label{appendix:ellip_shear}

When comparing SNPE modeling to automated forward modeling, we additionally compare parameter predictions for external shear and ellipticity in Figures \ref{external_shear_comp} and \ref{ellipticity_comp}. External shear predictions are generally consistent, with the exception of W2M J1042+1641, where, as discussed in Section \ref{subsection:image_positions}, the network confused an artifact for a point-source image. There are three notable outliers in the ellipticity predictions. The first two, DES J0530-3308 and W2M J1042+1641, are already identified as outliers due to their image position recovery, and are discussed in Section \ref{subsection:image_positions}. The third, J0029-3814, was not identified as an outlier in the image position recovery test. Note that, as can be seen in Figure 9 of \cite{ertl2023tdcosmo}, the inferred ellipticities from the two traditional modeling techniques also disagree with each other for this lens. This configuration may be generally challenging to model, especially given the lack of a visible source galaxy arc in the F814W band.   

\section{Full Covariance NPE Tests}
\label{appendix:full_covariance}

\begin{figure*}[hbt!]
\begin{nolinenumbers}
\begin{center}
\subfloat[Shifted test set]{\includegraphics[scale=0.35]{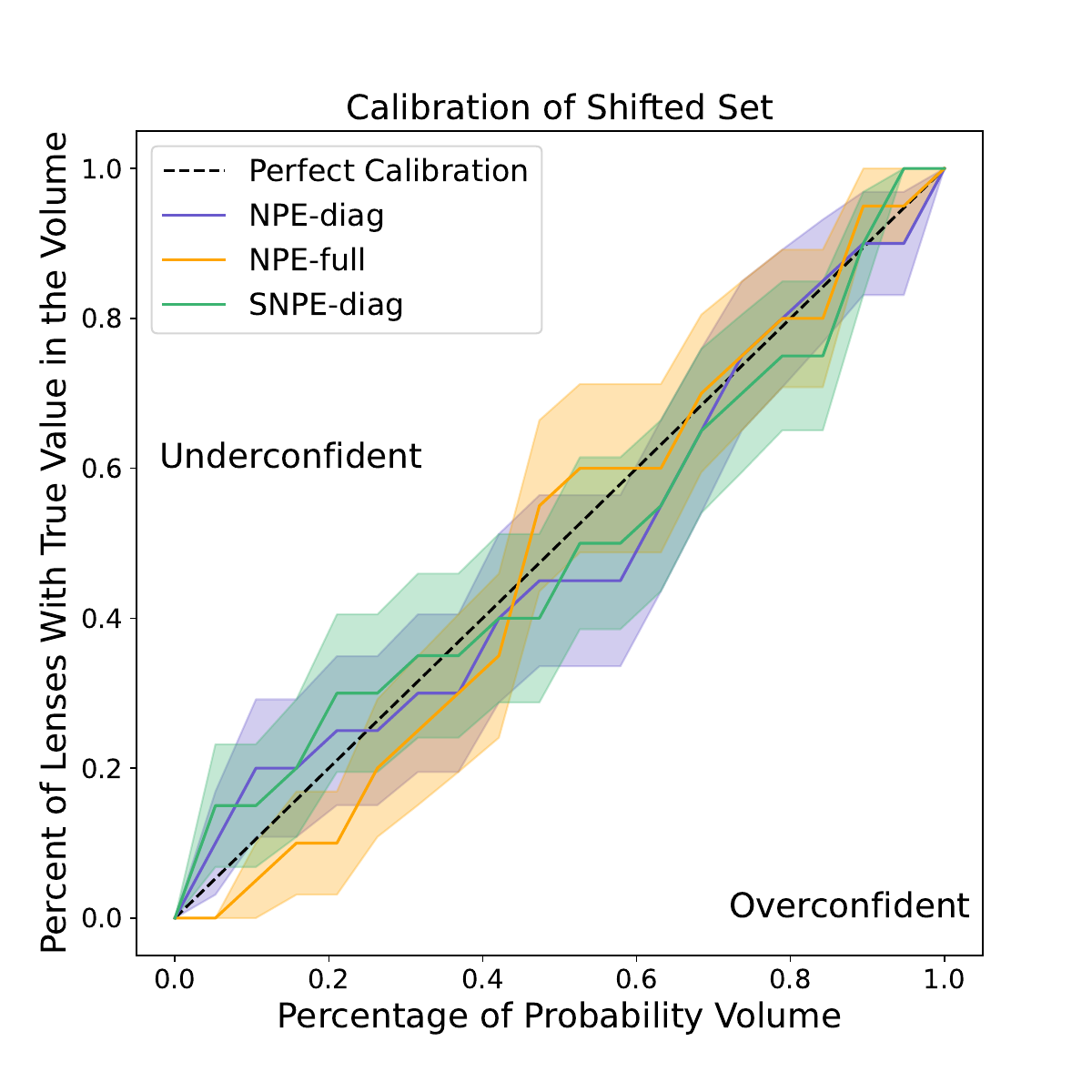}\label{shifted_calib_full}}
\quad
\subfloat[Doppelganger test set]{\includegraphics[scale=0.35]{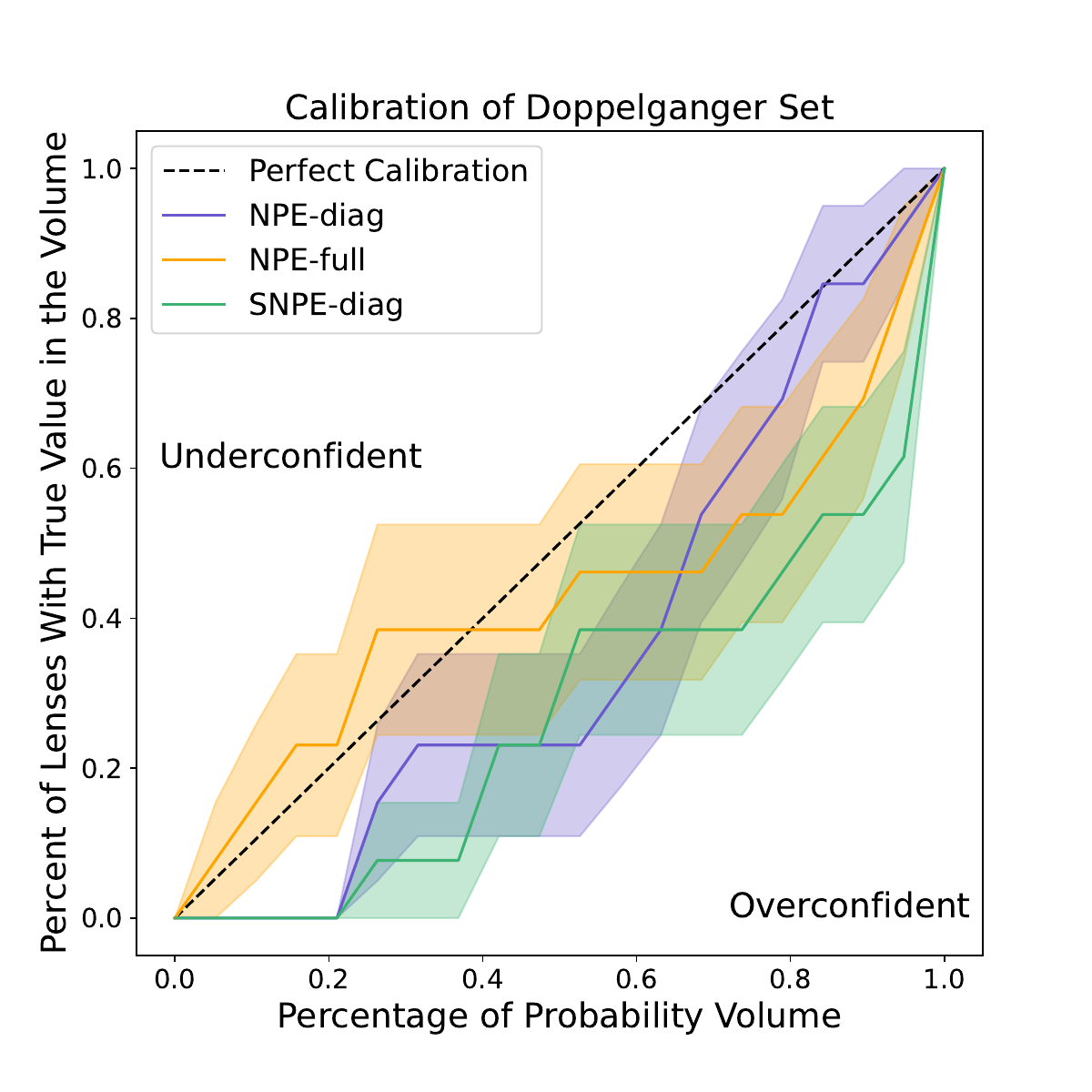}\label{doppel_calib_full}}
\caption{Calibration curves for the verification test sets with full covariance NPE included as a modeling option. Calibration from different methods is shown, with diagonal NPE (NPE-diag) shown in purple, full covariance NPE (NPE-full) shown in orange, and diagonal SNPE (SNPE-diag) shown in green. In perfectly calibrated posteriors (dashed line), a given x\% of the probability volume contains the truth x\% of the time. On the narrow test set, NPE-full calibration is consistent with the two other techniques. On the doppelganger test set, NPE-full calibration is closer to the perfect calibration curve, suggesting that inclusion of covariances in posteriors is crucial for the calibration of posteriors. }
\label{fig:fullcov_calibration_curves}
\end{center}
\end{nolinenumbers}
\end{figure*}

\begin{figure*}[hbt!]
\begin{nolinenumbers}
\begin{center}
\subfloat[Shifted test set]{\includegraphics[scale=0.27]{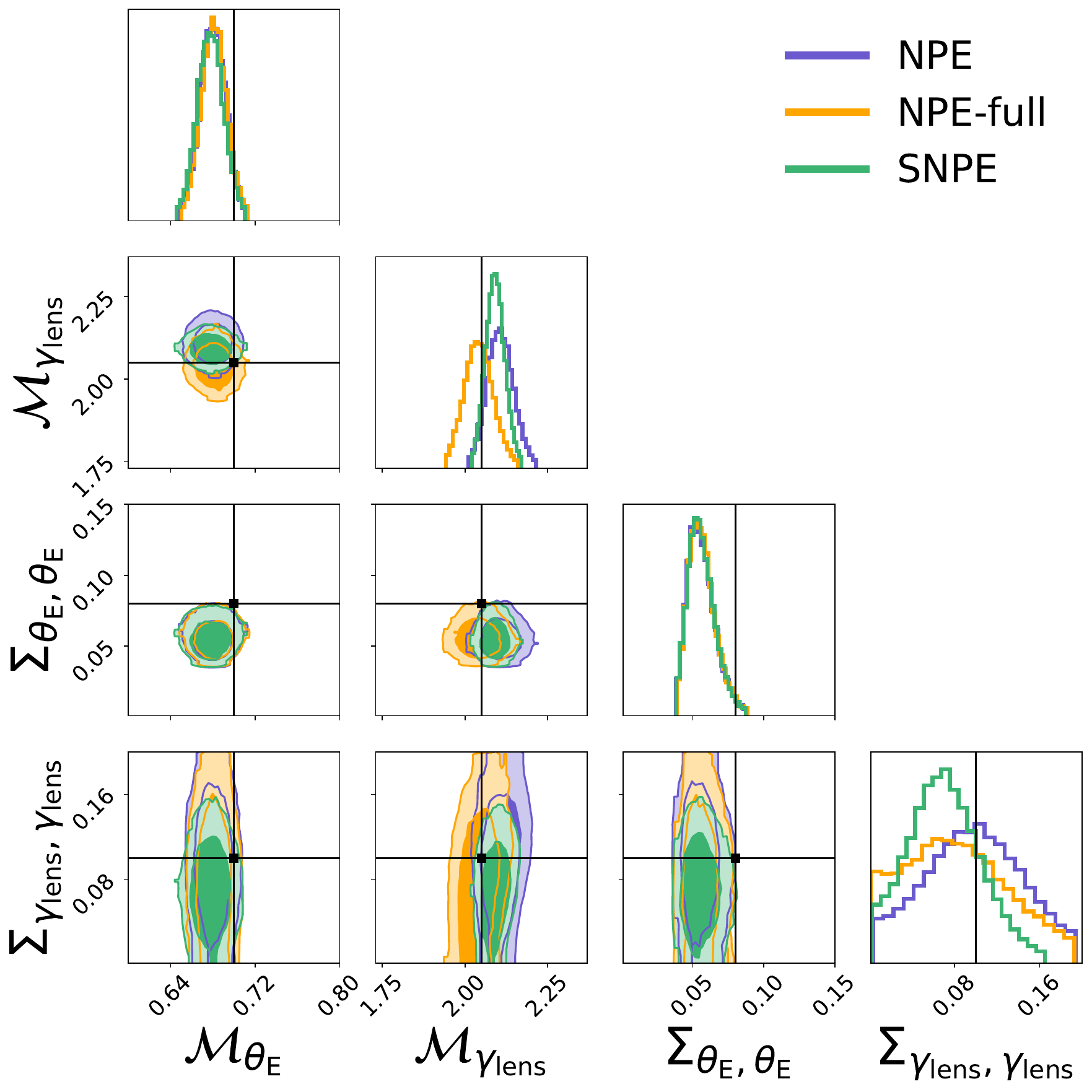}\label{shifted_HI_full}}
\quad
\subfloat[Doppelganger test set]{\includegraphics[scale=0.27]{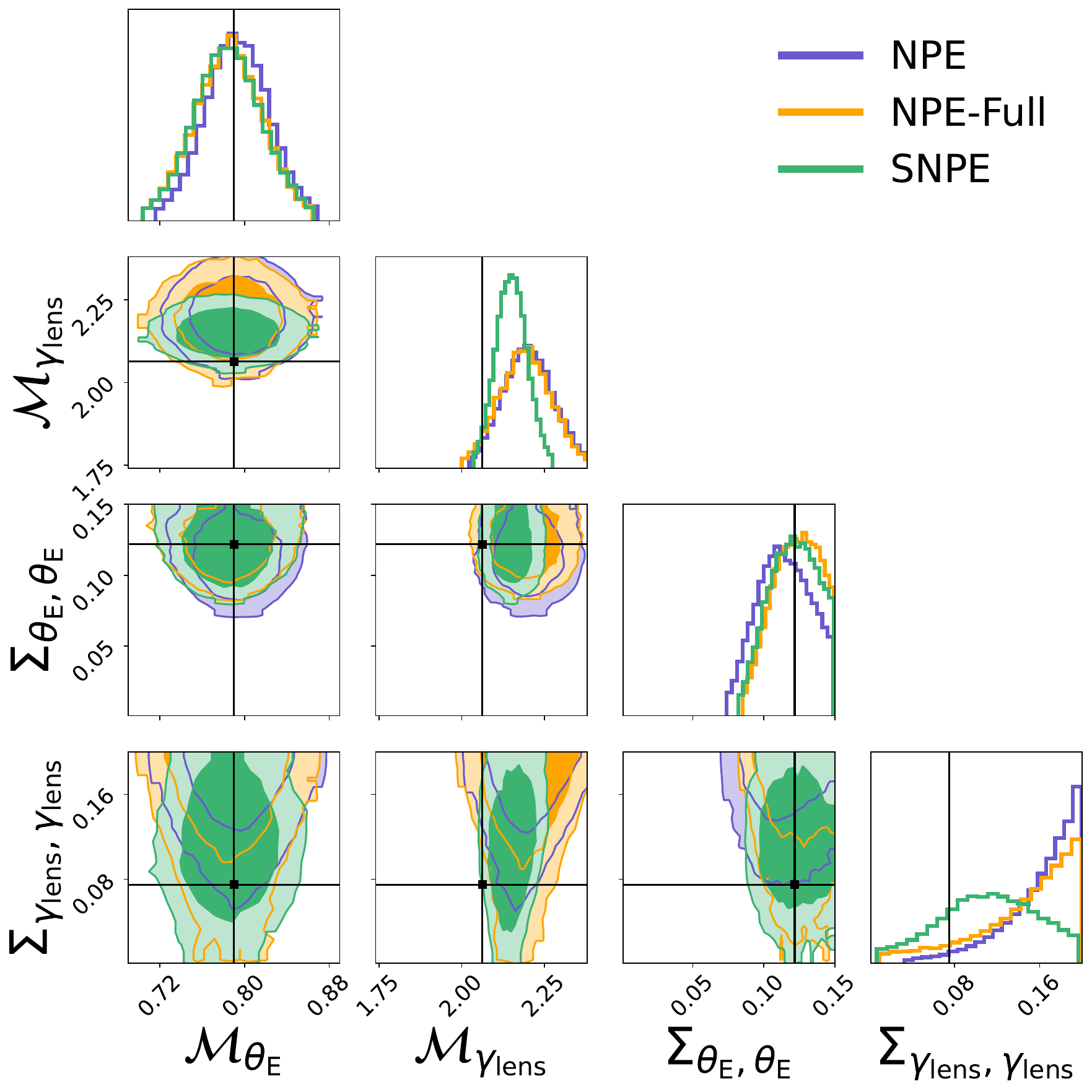}\label{doppel_HI_full}}
\caption{Two-dimensional contours of the hyperposterior p($\nu|\{d\}$) on the verification test sets. Inference from NPE with diagonal covariance posteriors is shown in purple. Inference from NPE with full covariance posteriors is shown in orange. Inference from SNPE with diagonal posteriors is shown in green. The ground truth is shown as a black line. Shaded contours are 68\% and 95\% intervals. Note for the doppelganger test set, we use the sample mean and standard deviation as a proxy ground truth.}
\label{fig:fullcov_npe_calibration_curves}
\end{center}
\end{nolinenumbers}
\end{figure*}

We investigate NPE with full covariance matrices. The inference technique remains the same as what is described in Section 
\ref{section:NPE}, except the final layer of the neural network now outputs 10 $\mu_k$ and the 45 elements of the log-cholesky decomposition of the full covariance matrix, $\Sigma_k$. We ran all of our analysis metrics on the narrow test set and doppelganger test set with this technique. We summarize performance with calibration plots in Figure \ref{fig:fullcov_calibration_curves} and hierarchical inference plots in Figure \ref{fig:fullcov_npe_calibration_curves}. On verification tests, the full covariance NPE method is well behaved. 

The next step was the application of full covariance NPE to the real HST images. In the first step of this analysis, we found indications that this method is not fully robust to complexities in the HST images. For four out of the 14 lenses (DES J0420$-$4037, PS J1606$-$2333, SDSS J0248+1913, W2M J1042+1641), we found that posterior PDFs had $\sigma_{k}(\gamma_{\text{lens}})$ larger than the $\sigma_{k}(\gamma_{\text{lens}})$ of the interim prior. For a fifth lens (DES J0530$-$3730), full covariance NPE results in a posterior wider than the prior in 3 dimensions ($\gamma_2$, $e_2$, $y_{\text{lens}}$). 

Ultimately, full covariance NPE shows promise given its well-behaved performance on the narrow test set and doppelganger test set. However, we hypothesize that full covariance NPE is more sensitive to details in the real HST data than the diagonal covariance NPE. Further work to increase the complexity of training simulations may be necessary before applying full covariance NPE on real data.

\section{Re-weighting Posteriors}
\label{appendix:reweighting}

Interim posteriors $q_\phi(\xi_{k} | d_{k}, \nu_{\text{int}})$ are influenced by the choice of hyperparameters $\nu_{\text{int}}$ that define the interim training prior. Once we have learned the true population hyperparameters via the hyperposterior $p(\nu | \{d\})$, we can use a re-weighting scheme to infer the individual lens posterior PDFs given the cPDF, which replaces the interim prior. This can also be thought of as using information from the population of lenses to refine the posterior of an individual lens. This re-weighting scheme is derived in \cite{Wagner_Carena_2021}, and ultimately we use the equation:

\begin{equation}
    q_\phi(\xi_{k} | d_{k}) \propto q_\phi(\xi_{k} | d_{k}, \nu_{\text{int}}) \int \frac{p(\xi_{k}|\nu)}{p(\xi_{k}|\nu_{\text{int}})} p(\nu|\{d\}_{\neq k}) d\nu
\label{reweighting_analytical}
\end{equation}

Note that to use this scheme, we need to perform a population-level inference k times, excluding one lens at a time to generate $p(\nu|\{d\}_{\neq k})$ for each lens. In practice, we compute the integral in Equation \ref{reweighting_analytical} using importance sampling: 

\begin{equation}
    q_\phi(\xi_{k} | d_{k}) \propto q_\phi(\xi_{k} | d_{k}, \nu_{\text{int}}) \frac{1}{N} \sum_{\nu \sim p(\nu|\{d\}_{\neq k})} \frac{p(\xi_{k}|\nu)}{p(\xi_{k}|\nu_{\text{int}})}
\label{reweighting_numerical}
\end{equation}

\begin{figure*}[hbt!]
\begin{nolinenumbers}
\begin{center}
\subfloat[Shifted test set]{\includegraphics[scale=0.35]{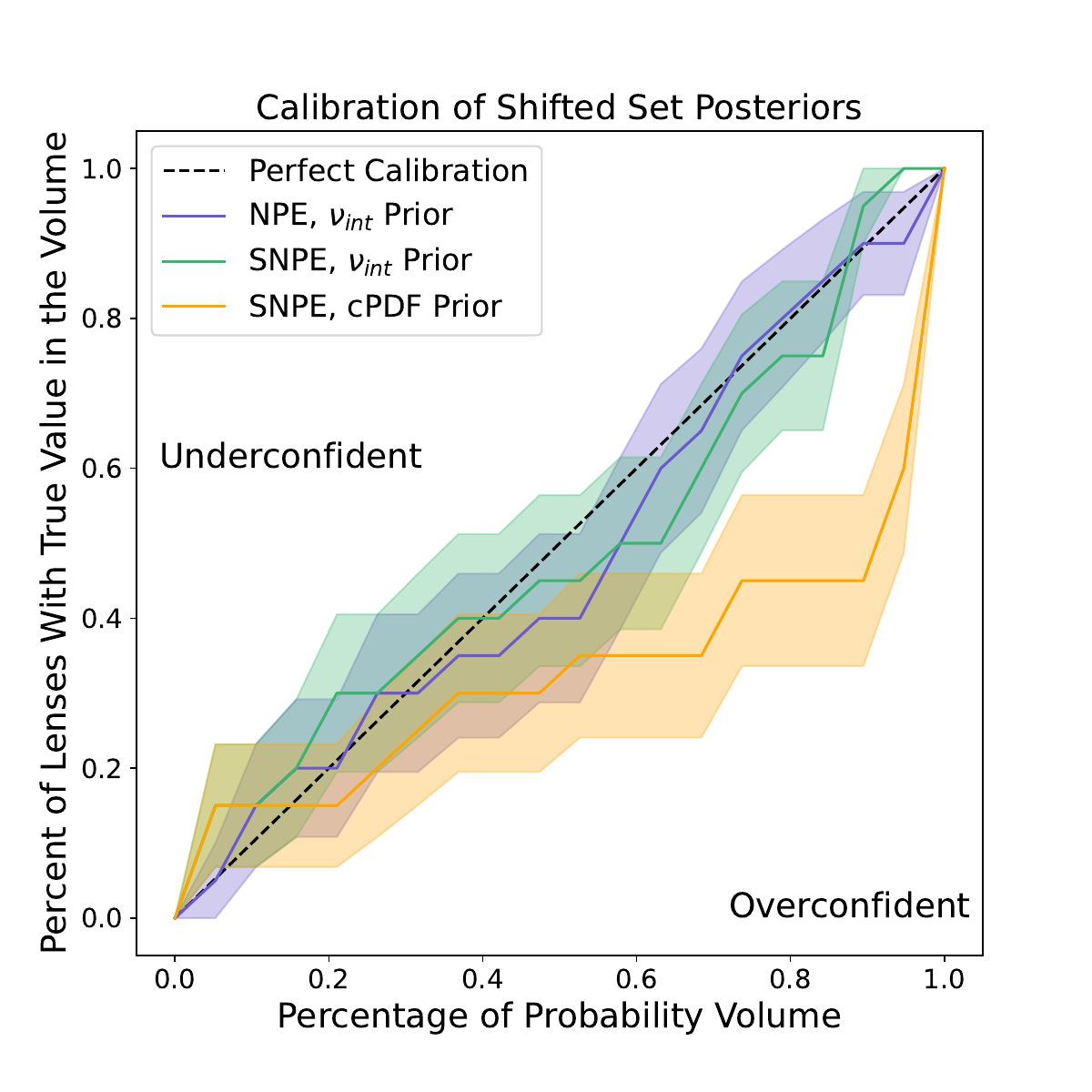}\label{rw_shifted_calib}}
\quad
\subfloat[Doppelganger test set]{\includegraphics[scale=0.35]{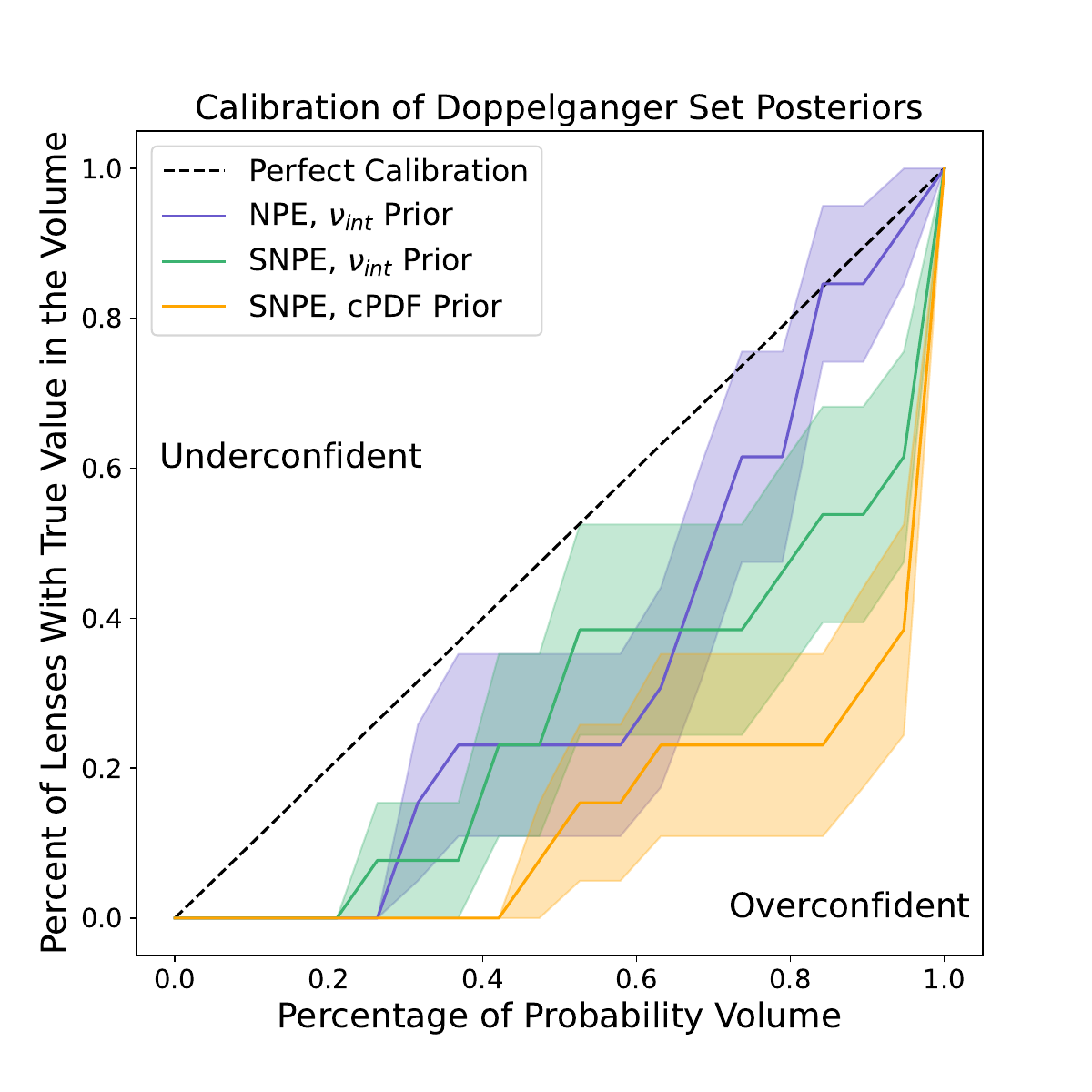}\label{rw_doppel_calib}}
\caption{Calibration curves for the verification test sets with re-weighting. In perfectly calibrated posteriors (dashed line), a given x\% of the probability volume contains the truth x\% of the time. Calibration of interim NPE posteriors is shown in purple. Calibration of interim SNPE posteriors is shown in green. Interim posteriors are generated under the informative prior $p(\xi_{k}|\nu_{\text{int}})$. Calibration of re-weighted SNPE posteriors is shown in yellow. Re-weighted SNPE posteriors effectively have a different prior assumption, with the cPDF being the new prior. The shaded region encompasses 1$\sigma$ uncertainty. Re-weighting makes error calibration worse, in the direction of increasing overconfidence, on both test sets.}
\label{reweighted_calibration_curves}
\end{center}
\end{nolinenumbers}
\end{figure*}

We applied this hierarchical re-weighting of individual posterior PDFs on the verification test sets. To re-weight the posteriors, we take samples from $q_\phi(\xi_{k} | d_{k}, \nu_{\text{int}})$, and assign each sample a weight by evaluating Equation \ref{reweighting_numerical}. The resulting calibration curves are shown in Figure \ref{reweighted_calibration_curves}. We find that the re-weighted posteriors have worse calibration, with the posterior widths becoming too small. We have three hypotheses for the cause of this effect. First, this could be due to imperfect inference of the individual posteriors. Any underlying bias in the initial posteriors influences the population model, and is then folded back in, potentially amplifying any smaller biases. A second hypothesis is that the re-weighting scheme is under-sampled, and probability density cannot move into the tails of a distribution when it needs to. The third hypothesis is that the functional form we asserted for the conditional PDF was insufficiently expressive, leading to artificially strong constraints on the individual lens model parameters. This third hypothesis only holds for the doppelganger test set, since the shifted test set distribution is Gaussian by definition.

\end{document}